\documentclass{aa}
\usepackage{txfonts}
\usepackage{natbib}
 \usepackage{times}
 \usepackage{amsfonts}
 \usepackage{amssymb}
 \usepackage{graphicx}
 \usepackage{aalongtable}

\begin{document}
\title{A PDR-Code Comparison Study}
\author{
M. R\"ollig\inst{1,4},
N. P. Abel\inst{2},
T. Bell\inst{3,17},
F. Bensch\inst{1},
J. Black\inst{15},
G. J. Ferland\inst{2},
B. Jonkheid\inst{5},
I. Kamp\inst{6},
M.J. Kaufman\inst{7},
J. Le Bourlot\inst{8},
F. Le Petit\inst{8,15},
R. Meijerink\inst{5},
O. Morata\inst{16},
V. Ossenkopf\inst{4,10},
E. Roueff\inst{8}, 
G. Shaw\inst{2},
M. Spaans\inst{9}, 
A. Sternberg\inst{11},
J. Stutzki\inst{4} , 
W.-F. Thi\inst{12}, 
E. F. van Dishoeck\inst{5}, 
P. A. M. van Hoof\inst{13},
S. Viti\inst{3},
M.G. Wolfire\inst{14}
}
\institute{
Argelander-Institut f\"ur Astronomie \thanks{Founded by merging of the
Sternwarte, Radiastronomisches Institut and Institut f\"ur Astrophysik und 
Extraterestrische Forschung}, Universit\"at Bonn, Auf dem H\"ugel 71, D-53121 Bonn, Germany
\and University of Kentucky, Department of Physics and Astronomy, Lexington, KY 40506, USA
\and Department of Physics \& Astronomy, University College London, Gower Street, London WC1E 6BT
\and I. Physikalisches Institut, Universit\"at zu K\"oln, Z\"ulpicher Str. 77, D-50937 K\"oln,
Germany
\and Leiden Observatory, P.O. Box 9513, NL-2300 RA Leiden, Netherlands
\and Space Telescope Science Division of ESA, Space Telescope Science Institute, Baltimore, MD 21218, USA
\and Department of Physics, San Jose State University, 1 Washington Square, San Jose, CA 95192, USA
\and LUTH UMR 8102, CNRS and Observatoire de Paris, Place J. Janssen 92195 Meudon Cedex, France
\and Kapteyn Astronomical Institute, PO Box 800, 9700 AV Groningen, The Netherlands
\and SRON National Institute for Space Research, Postbus 800, 9700 AV Groningen, The Netherlands
\and School of Physics and Astronomy, Tel Aviv University, Ramat Aviv 69978, Israel
\and Institute for Astronomy, The University of Edinburgh, Royal Observatory, Blackford Hill, Edinburgh EH9 3HJ, U.K.
\and Royal Observatory of Belgium, Av. Circulaire, 3 - Ringlaan 3,1180 BRUXELLES - BRUSSEL 
\and Astronomy Department, University of Maryland, College Park, MD 20742-2421, USA
\and Onsala Space Observatory, 439 92 Onsala, Sweden
\and LAEFF, Villafranca del Castillo, Apdo. 50727, E-28080 Madrid, Spain
\and California Institute of Technology, 1200 E. California Blvd, Pasadena CA 91125, USA
}
\authorrunning{R\"ollig et al.}

\offprints{M. R\"ollig,\\ \email{roellig@ph1.uni-koeln.de}}

\abstract{}
{We present a comparison between independent computer codes, modeling the physics and chemistry of
interstellar photon dominated regions (PDRs).  Our goal was to understand
the mutual differences in the PDR codes and their effects on the physical and chemical structure
of the model clouds, and to converge the output of different codes to a
common solution.}
{ A number of benchmark models have been created, covering low  and high  gas
densities $n=10^3,10^{5.5}$~cm$^{-3}$ and far ultraviolet intensities $\chi=10,10^5$ in units of the Draine field (FUV: $6<\,h\,\nu\,<\,13.6$ eV). The benchmark models were 
computed in two ways: one set assuming constant temperatures, thus testing the consistency of the chemical network and 
photo-processes, and a second set determining the temperature self consistently by
solving the thermal balance, thus testing the modeling of the heating and cooling mechanisms
 accounting for the detailed energy balance throughout the clouds.}
{We investigated the impact of PDR geometry and agreed on the comparison
of results from spherical and plane-parallel PDR models. We 
identified a number of key processes 
governing the chemical network
which have been treated differently in the various codes such as
 the effect of PAHs on the electron density or the temperature dependence of 
the dissociation of CO by cosmic ray induced secondary photons,
and defined a proper common treatment.
We established a comprehensive set of reference 
models for ongoing and future PDR model bench-marking and were able to increase the 
agreement in model predictions for all benchmark models significantly. 
Nevertheless, the remaining spread in the computed observables such
as the atomic fine-structure line intensities serves as a warning that 
there is still a considerable uncertainty when interpreting
astronomical data with our models.}{}

\maketitle
\section{Introduction}
Interstellar photon dominated regions or photodissociation regions (PDRs) play 
an important role in modern astrophysics as they are responsible for many
emission characteristics of the ISM, and dominate the infrared 
and sub-millimetre spectra of star formation regions and galaxies as a whole.
Theoretical models addressing the 
structure of PDRs have been available for approximately 30 
years and have evolved into advanced computer codes
accounting for a growing number of physical effects with increasing accuracy.
These codes have been developed
with different goals in mind: some are geared
to efficiently model a particular type of region, e.g. HII regions, protoplanetary disks, planetary nebulae, 
 diffuse clouds, etc.;
others emphasize a strict handling of the micro-physical  processes in full detail
(e.g. wavelength dependent absorption), but at the cost of increased computing time. Yet others aim 
at efficient and rapid calculation of large model grids for comparison with observational
data, which comes at the cost of pragmatic approximations using effective rates rather than detailed treatment.
As a result, the
different models have focused on the detailed simulation of particular processes determining the structure in the 
main regions of interest while using only rough approximations for other processes. 
The model setups vary strongly among different model codes. This includes
the assumed model geometry, their physical and chemical structure, the choice of free 
parameters, and other details. Consequently 
it is not always straightforward to directly compare the results from different PDR
codes. Taking into account that there are multiple ways of implementing physical 
effects in numerical codes, it is obvious that the model output of different PDR codes
can differ from each other. As a result, significant variations in the
physical and chemical PDR structure 
predicted by the
various PDR codes can occur.
This divergence would prevent a unique interpretation of observed data in terms of the
parameters of the observed clouds.
Several new facilities such as Herschel, SOFIA, APEX, ALMA, and others
will become available over the next years and
will deliver many high quality observations of line and dust continuum emission in the sub-millimeter
and FIR
wavelength regime. Many important PDR tracers emit in this range ([CII] (158$\mu$m), [OI] (63 and 146 $\mu$m), 
[CI] (370 and 610 $\mu$m), CO (650, 520, ..., 57.8 $\mu$m), H$_2$O, etc.). In order to reliably 
analyze these data we need a set of high quality tools, including PDR models
that are well understood and properly debugged.
As an important preparatory step toward these missions an international cooperation 
between many PDR model groups was initiated. The goals 
of this PDR-benchmarking were:

\begin{itemize}
\item to understand the differences in the different code results
\item to obtain (as much as possible) the same model output with every PDR code when using the same input
\item to agree on the correct handling of important processes
\item to identify the specific limits of applicability of the available codes
\end{itemize}

To this end, a PDR-benchmarking workshop was held at the Lorentz Center in 
Leiden, Netherlands in 2004 to jointly work on these topics 
\footnote{URL: {\tt http://www.lorentzcenter.nl/}}. In this paper we present
the results
from this workshop and the results originating from the follow-up activities.
A related workshop to test line radiative transfer codes was held in 1999
\citep[see][]{vZ02}. 

It is not the purpose
of the benchmarking to present a preferred solution or a preferred code. PDRs are found
in a large variety of objects and under very different conditions. To this end,
it was neither possible nor desirable to develop a {\it generic} PDR code, able to model every 
possible PDR. Furthermore, the benchmarking
is not meant to model any 'real' astronomical object. The main purpose of this
study is technical not physical. This is also reflected in the choice of the adopted
incomplete chemical reaction network (see \S~\ref{benchmark_models}).   
   
 In \S~\ref{physics} we briefly introduce the 
physics involved in PDRs, in \S~\ref{modelling} we introduce some key features
in PDR modeling. \S~\ref{benchmark_models} describes the setup of the benchmark calculations
and \S~\ref{results} presents the results for a selection of benchmark calculations and
gives a short review over the participating codes.  In \S~\ref{summary} we discuss the results and 
summarize the lessons learned from the benchmark effort.
A tabular overview of the individual code characteristics is given in the Appendix.

\section{The Physics of PDRs}\label{physics}
PDRs are
traditionally defined as regions where H$_2$-non-ionizing far-ultraviolet
photons from stellar sources control the gas heating and chemistry.
Any ionizing radiation is assumed to be absorbed in the narrow
ionization fronts located between adjacent HII regions and the 
PDRs\footnote{This distinction is clearer when referring
to PDRs as Photo-Dissociation Regions, since  molecules
are hardly found in HII regions}.
In PDRs the gas  is heated by the far-ultraviolet radiation (FUV, $6\,<\,h\nu\,<\, 13.6$ eV, 
from the ambient UV field and from hot stars) and cooled via the emission
of spectral line radiation of atomic and molecular species  and continuum emission by dust
 (Hollenbach \& Tielens 1999, Sternberg 2004). The
FUV photons heat the gas by means of photoelectric emission from grain surfaces and
polycyclic aromatic hydrocarbons (PAHs) and
by collisional de-excitation of vibrationally excited H$_2$ molecules. Additional contributions
to the total gas heating comes from H$_2$ formation, dissociation of H$_2$, dust-gas
collisions in case of dust temperatures exceeding the gas temperature,
 cosmic ray heating, turbulence, and from chemical heating. 
At low visual extinction $A_\mathrm{V}$ into the cloud/PDR the gas is cooled  by emission of atomic fine-structure lines, mainly [OI] 63$\mu$m and [CII] 158$\mu$m. At larger depths, millimeter, sub-millimeter and far-infrared
molecular rotational-line cooling (CO, OH, H$_2$, H$_2$O) becomes important, and a correct treatment
of the radiative transfer in the line cooling is critical. The balance between heating 
and cooling determines the local gas temperature. The local FUV intensity also influences the
 chemical structure, i.e. the abundance of the individual chemical constituents of the
gas. The surface of PDRs is mainly dominated by
reactions induced by UV photons, especially the ionization and dissociation of atoms and molecules. 
With diminishing FUV intensity at higher optical depths more complex species may be formed without being radiatively
destroyed immediately. Thus the overall structure of a PDR is the result of a complex interplay
between radiative transfer, energy balance, and chemical reactions. 
   
\section{Modeling of PDRs}\label{modelling}
The history of PDR modeling dates back to the early 1970's \citep{HS71b, jura74, glassgold75, black77}
with steady state models for the transitions from H to H$_2$ and from C$^+$ to CO. In the 
following years a number of models, addressing the chemical and thermal structure
of clouds subject to an incident flux of FUV photons have been developed \citep{dejong80,TH85,vDB88, SD89,
HTT91,lebourlot93, stoerzer96}. Additionally, a number of models, focusing on certain
aspects of PDR physics and chemistry were developed, e.g. models accounting for 
time-dependent chemical networks, models of clumped media, and turbulent PDR models
\citep{hill78, wagenblast88, deboisanger92, bertoldi96, lee96, hegmann96, spaans96, nejad99, roellig02, 
bell05}.  Standard PDR models generally do 
not account for dynamical properties
of gas but there are some studies that consider the advection problem rather
than the steady state approach \citep[e.g.][]{stoerzer88}. For a more detailed review 
see \cite{HT99}. 

In order to numerically model a PDR it is necessary to compute all  
local properties of a cloud 
such as the relative abundances of the gas constituents
together with their level populations, temperature of gas and dust, gas pressure, 
composition of dust/PAHs, and many more. This local treatment is  
complicated by the radiation field which couples remote parts of the cloud. 
The local mean radiation field, which is responsible for photochemical reactions, 
gas/dust heating, and excitation of molecules depends on the position 
in the cloud
and the (wavelength dependent) absorption along the lines of sight toward this position.
 This non-local coupling makes numerical PDR calculations
a CPU time consuming task.

PDR modelers and observers approach the PDRs from opposite sides: PDR models start by calculating
the local properties of the clouds such as the local CO density and the corresponding gas temperature
and use these local properties to infer the expected global properties of the cloud like total 
emergent emissivities or fluxes and column densities. The observer on the other hand starts by
observing global features of a source and tries to infer the local properties from that. The 
connection between local and global properties is complex and not necessarily unambiguous.
Large variations e.g. in the CO density at the 
surface of the cloud may hardly affect the overall 
CO column density due to the dominance of the
central part of the cloud with a high 
 density. If one is interested in the total column density it does not matter whether
 different codes produce a different
surface CO density. For the interpretation of high-J CO emission lines, 
however, different CO densities in the
outer cloud layers make a huge difference since high 
temperatures are required to produce 
high-J CO fluxes.
Thus, if different PDR model codes deviate in their predicted cloud structures, this 
may affect the interpretation
of observations and may prevent inference of the 'true' structure behind the observed data.
To this end it is very important to understand the origin of present differences
in PDR model calculations. Otherwise it is impossible to rule out alternative 
interpretations. The ideal situation,
from the modelers point of view, would be a complete knowledge of the true 
local structure of a real cloud {\bf and} their 
global observable properties. This would easily allow us to calibrate PDR models.
Since this case is unobtainable, we take one step back and apply a different approach: If 
all PDR model codes use
exactly the same input and the same model assumptions they 
should produce the same predictions.
   
Because of the close interaction between chemical and thermal balance and
radiative transfer, PDR codes typically iterate through the following computation steps:
1) solve the local chemical balance to determine local densities, 
2) solve the local energy balance to estimate the local physical properties like temperatures, 
pressures, and level populations, 
3) solve the radiative transfer,
4) for finite models it is
necessary to successively iterate steps 1)-3).
Each step requires
 a variety of assumptions and simplifications. Each
of these aspects can be investigated to great detail and complexity (see for example \cite{vZ02} for a
discussion of NLTE radiative transfer methods), but
the explicit aim of the PDR comparison workshop was to understand the interaction of 
all computation steps mentioned above.
 Even so it was necessary 
to considerably reduce the model complexity in order to disentangle cause and effect.

\subsection{Description of Sensitivities and Pitfalls}
Several aspects of PDR modeling have shown the need for detailed discussion, easily
resulting in misleading conclusions if not treated properly:
\subsubsection{Model Geometry}
\begin{figure}[hbt]
\begin{center}
\resizebox{\hsize}{!}{\includegraphics{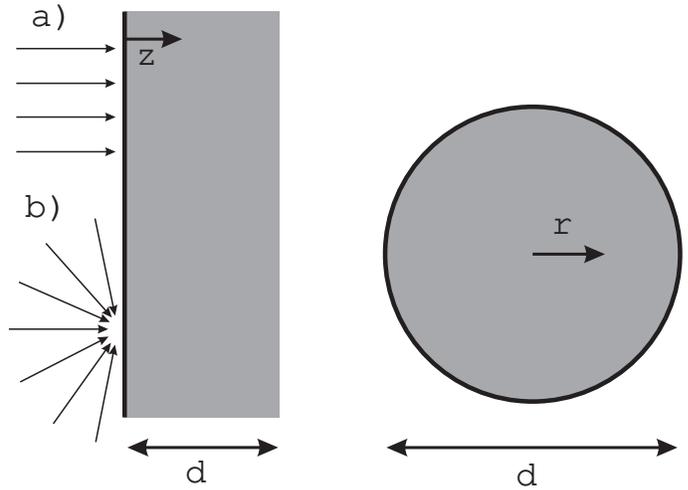}}
\caption{Common geometrical setups of a model PDR. The surface of any
plane-parallel or spherical cloud is illuminated either a) uni-directional or 
b) isotropically.}\label{geom}
\end{center}
\end{figure}
\begin{figure}[htb]
\begin{center}
\resizebox{\hsize}{!}{\includegraphics{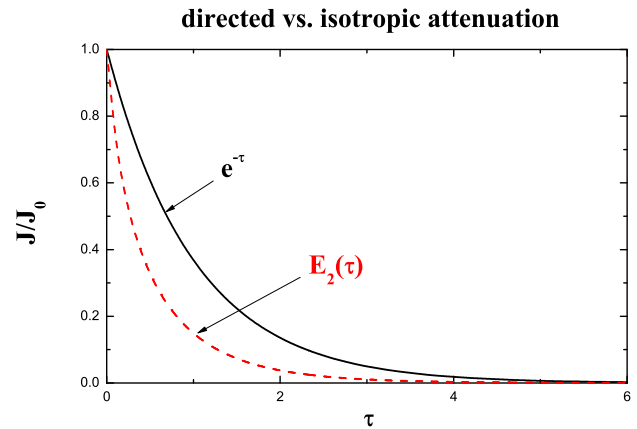}}
\caption{Comparison of attenuation of the mean intensity for the case of an 
uni-directional and isotropically illuminated medium.
The solid line gives the attenuation due to uni-directional illumination, while
the dashed line gives the attenuation for an isotropic FUV radiation where $\tau$ means the 
optical depth perpendicular to the surface of the cloud.}\label{exp}
\end{center}
\end{figure}
Two common geometrical setups of model PDRs are shown in Figure \ref{geom}. Most
PDR models feature a plane-parallel geometry, illuminated either from one side or from both sides. This geometry
naturally suggests a directed illumination, perpendicular to the cloud surface. This simplifies the
radiative transfer problem significantly, since it is sufficient to account for just one 
line of sight, if we ignore scattering out of the line of sight \citep{flannery80}. Since 
most plane-parallel PDR models are infinite perpendicular to the 
cloud depth $z$ 
it is also straightforward to account for an isotropic FUV irradiation
 within the pure 1-D formalism. 
For a spherical geometry one can exploit the model symmetry only for 
a FUV field isotropically impinging onto the cloud. In finite plane-parallel 
and spherical models iterations over the depth/radial structure are mandatory
because radiation is coming from multiple directions, passing 
through cloud elements 
for which the physical and chemical structure and hence opacities 
have not been calculated in the same iteration step. 
To account for this 'backside' illumination it is essential to iterate 
on the radiation field.  
 
The most important quantity describing the radiation field in PDR models
is the local mean intensity (or alternatively the energy density) as given by:
\begin{equation}\label{meanj}
J_\nu=\frac{1}{4\,\pi}\int\,I_\nu\, d\Omega\,\,\,[\mathrm{erg\; cm^{-2}\; s^{-1}\; Hz^{-1}\; sr^{-1}}]
\end{equation}
with the specific intensity $I_\nu$ being averaged over the solid angle $\Omega$. Note that 
when referring to the ambient FUV in units 
of Draine $\chi$ \citep{draine78} or Habing $G_0$ \citep{habing68} fields, these are always given as averaged over
$4\pi$. If we place a model cloud of sufficient optical thickness in such
an average FUV field, the resulting local mean intensity at the cloud edge is half the value of that without the cloud.
 
\begin{figure}[htb]
\begin{center}
\resizebox{\hsize}{!}{\includegraphics{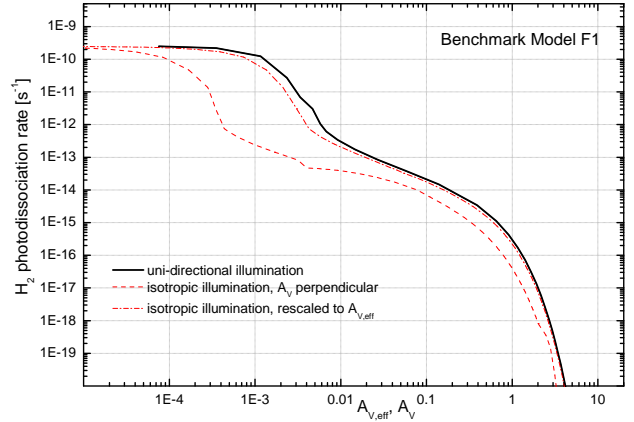}}
\caption{H$_2$ photo-dissociation rates resulting from uni-directional FUV illuminated clouds compared to an isotropic
illumination. The results from isotropic models are plotted vs. the perpendicular $A_\mathrm{V}$ and vs. $A_\mathrm{V,eff}$. }\label{avtoaveff}
\end{center}
\end{figure}

The choice between directed and isotropic FUV fields directly influences the attenuation due to dust. In the
uni-directional case the FUV intensity along the line of sight is attenuated according to $\exp(-\tau_\nu)$, where $\tau_\nu$ is the
 optical depth of the dust at frequency $\nu$. For pure absorption the radiative 
transfer equation  becomes:
\begin{equation}\label{radtrans}
\mu\,\frac{d I_\nu(\mu,x)}{dx}=-\kappa_\nu\,I_\nu(\mu,x)\;\;\;.
\end{equation}
with the cosine of the radiation direction $\mu=\cos\Theta$, the cloud depth $x$, and the absorption coefficient
 $\kappa_\nu$, with the simple solution  $J_\nu/J_{\nu,0}=\exp(-\tau_\nu\,\mu)$ for
a semi-infinite cloud. For the isotropic case,  $I_{\nu,0}(\mu)=J_{\nu,0}=const.$, 
integration of Eq. \ref{radtrans} leads to the second order exponential integral:
\begin{equation}
J_\nu/J_{\nu,0}=E_2(\tau_\nu)=\int_0^1\,\frac{\exp(-\tau_\nu\,\mu)}{\mu^2}\,d\mu 
\end{equation}
As seen in Figure \ref{exp} the attenuation with depth in the isotropic case is 
significantly different from the uni-directional case. A common way to describe the depth dependence of
a particular quantity in PDRs is to plot it against $A_\mathrm{V}$, which is a direct measure of the traversed column of attenuating
material.  In order to 
compare the uni-directional and the isotropic case
 it is necessary to rescale them to the same axis. It is possible to define an effective
 $A_\mathrm{V,eff}=-\ln[E_2(A_\mathrm{V}\,k)]/k$ with $k=\tau_\mathrm{UV}/A_\mathrm{V}$ in the 
isotropic case, where $A_\mathrm{V}$ is the attenuation perpendicular to the surface and UV is in the range
$6<h\nu<13.6$. In this paper all results from spherical models are scaled to  $A_\mathrm{V,eff}$. Figure 
\ref{avtoaveff} demonstrates the importance of scaling results to 
an appropriate $A_\mathrm{V}$ scale. It shows the local H$_2$ photo-dissociation rate for two different FUV illumination 
geometries. The solid line represents a standard uni-directional illumination perpendicular to the cloud surface as
given in many standard plane-parallel PDR codes. The dashed line is the result from an isotropic illumination 
plotted against the standard 'perpendicular' $A_\mathrm{V}$. The offset to the uni-directional case is significant. After rescaling
to an appropriate $A_\mathrm{V,eff}$ both model results are in good agreement. Please note, that in general it is not possible to achieve perfect 
agreement as there is a spectrum involved with a spread 
of $k$ values across the UV. 

The attenuation of FUV radiation is additionally complicated if we account for dust scattering.
 For a full treatment by Legendre polynomials see \citet{flannery80}.
In case of small scattering angles $g=\langle \cos \theta \rangle \approx 1$
 the scattering can be approximated by an effective 
forward attenuation $\tau(1-\omega)$, where $\omega$ is the scattering albedo. Thus, more material is needed
to obtain the same attenuation as in the case without scattering. Hence a proper 
scaling of $A_\mathrm{V}$ is necessary. In case of clumped gas this becomes even more complex. The
presence of stochastic density fluctuations leads to a substantial reduction of the
effective optical depth as demonstrated by \citet{hegmann03}. All this has to 
be considered when calculating the photodissociation and 
photoionization rates, when the attenutation with depth is represented by
simple exponential forms,
$\exp(-k_i\,A_\mathrm{V})$ \citep[e.g.][]{VD88,roberge91}, where the factor $k_i$ accounts for the 
wavelength dependence of the photoprocess $i$\footnote{In this context the term photoprocess refers to either 
photodissociation or photoionization.}.

\subsubsection{Chemistry}\label{chemistry}
PDR chemistry has been addressed in detail by many authors \citep{TH85,vDB88,HTT91,fuente93,lebourlot93,jansen95,SD95,lee96,bakes98,walmsley99,savage04,teyssier04,fuente05,meijerink05}.
These authors discuss numerous aspects of PDR chemistry in great detail and  give a comprehensive overview of the field. ´Here we repeat some crucial points in the chemistry of PDRs in order to motivate the benchmark standardization and 
to prepare the discussion of the benchmark result. 
   
In PDRs photoprocesses are very important due to the high FUV intensity, as 
well as reactions with abundant hydrogen atoms. The formation and destruction 
of  H$_2$, heavily influenced by the FUV field, is of major importance for 
the chemistry in PDRs. H$_2$ forms on grain surfaces, a 
process which crucially depends on the temperatures of the gas and 
the grains \citep{HS71a, cazaux04}, which themselves depend on the 
local cooling and heating, governed by the FUV. The photo-dissociation 
of H$_2$ is a line absorption process and, thus is subject to effective 
shielding \citep{vDB88}. This leads to a sharp transition from atomic
 to molecular hydrogen once the H$_2$ absorption lines are optically 
thick. The photo-dissociation of CO is also a line absorption process, 
additionally complicated by the fact that the broad H$_2$ absorption 
lines overlap with CO absorption lines. Similar to  H$_2$ this leads 
to a transition from atomic carbon to CO.  For $A_\mathrm{V}<1$ carbon 
is predominantly present in ionized form. For an assumed FUV field of
 $\chi=1$, CO is formed at about $A_\mathrm{V}\approx 2$. This results 
in the typical PDR stratification of H/~H$_2$ and C$^+$/~C/~CO. The 
depth of this transition
zone depends on the physical parameters but also on the contents of the 
chemical network: for example the inclusion of PAHs into the chemical 
balance calculations shifts the C$^+$ to C transition to smaller
 $A_\mathrm{V,eff}$ \citep[e.g.][]{lepp88,bakes98}.

The solution of the chemical network itself covers the destruction and formation reactions of all chemical species considered. For each included
species $i$ this results in a balance equation of the form:
\begin{eqnarray}\label{chemrateeq}
\frac{d n_i}{d t}&=&\sum_j\sum_k\,n_j\,n_k\,R_{jki}\,+\,\sum_l\,n_l\,\zeta_{li}\,\nonumber \\
& &-\,n_i\left(\sum_l\zeta_{il}\,+\,\sum_l\sum_j\,n_j\,R_{ijl}\right)\label{chembal}
\end{eqnarray}
 Here $n_i$ denotes the density of species $i$.
The first two terms cover all formation processes while the last two terms account for all destruction reactions. $R_{jki}$ is the reaction 
rate coefficient for the reaction ${\mathrm X}_j\,+\,{\mathrm X}_k\,\to\,{\mathrm X}_i\,+\,...$ (X stands for species X), $\zeta_{il}$ is the local photo-destruction rate coefficient for ionization or dissociation of species $\mathrm{X}_i\,+h\,\nu\,\to {\mathrm X}_l +...$, either by FUV photons or by cosmic ray (CR) induced photons, and $\zeta_{li}$ is the local formation rate coefficient for formation of $\mathrm{X}_i$ by photo-destruction of species $\mathrm{X}_l$. For a stationary solution one assumes
$dn_i/dt=0$, while non-stationary models solve the differential equation (\ref{chemrateeq}) in time. 
The chemical network is a highly non-linear system of equations. Hence it is not self-evident that a unique
solution exists at all, multiple solution may be possible as demonstrated e.g. by \citet{lebourlot93} and \citet{boger06}.
\begin{figure*}
\centering\includegraphics[width=17 cm]{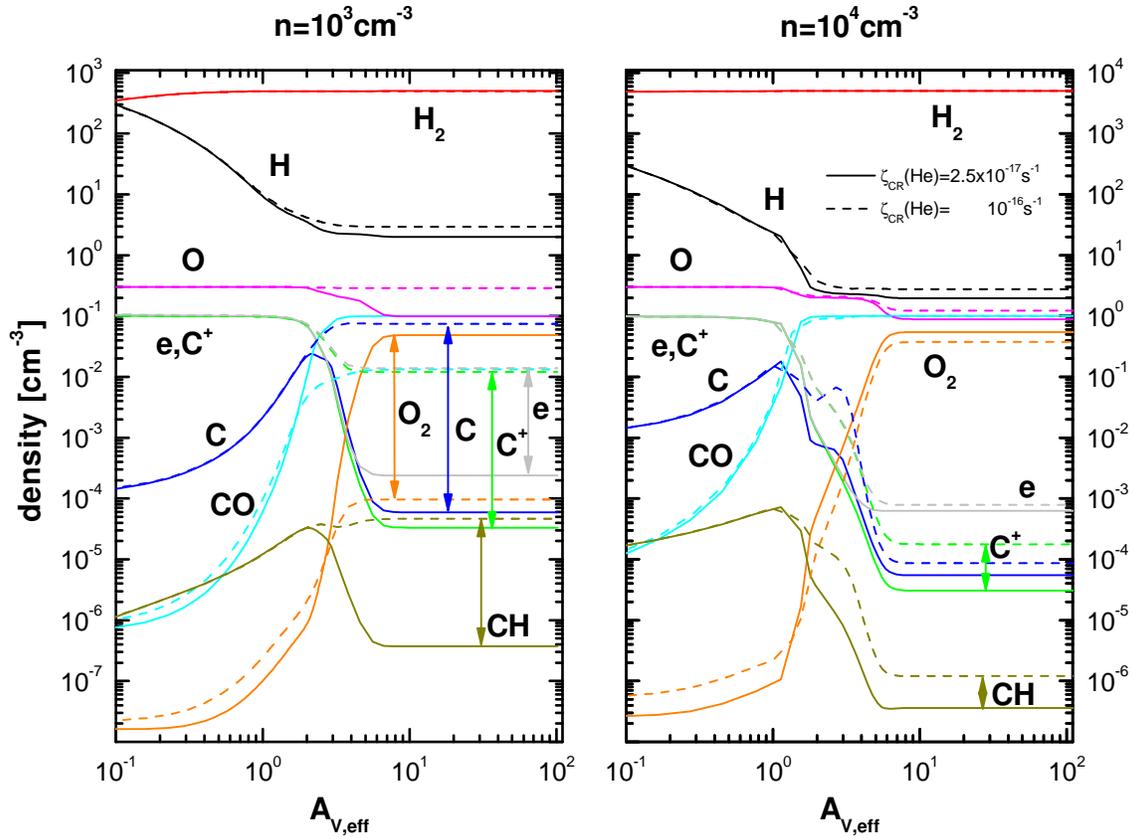}
\caption{The influence of the cosmic ray ionization rate 
on the chemical structure of a model cloud. The left panel shows results for 
Model F1 (n=10$^{3}$~cm$^{-3}$, $\chi=10$), the right panel gives results for
10 times higher densities (n=10$^{4}$~cm$^{-3}$, $\chi=10$).
 The solid lines give the results for a cosmic ray  
ionization rate of Helium, enhanced by a factor 4, the dashed lines are 
for the lower ionization rate.
The different colors denote different chemical species. The most prominent 
differences are highlighted with  colored arrows.}\label{F1-CRP}
\end{figure*}
 
They showed that bistability may occur
in gas-phase models (neglecting dust chemistry) of interstellar dark clouds 
in a narrow parameter range of approximately 
 $10^{3}$~cm$^{-3}\gtrsim\,n/\zeta_{-17}\gtrsim 10^{2}$~cm$^{-3}$ with 
the cosmic-ray ionization rate of molecular hydrogen
 $\zeta_\mathrm{CR}\equiv10^{-17}\zeta_{-17}$~s$^{-1}$. Within this range 
the model results may depend very sensitively on variations of input parameters
 such as $\zeta_\mathrm{CR}$ or the H$_3^+$ dissociative recombination rate. To 
demonstrate this we show the influence of varying ionization rates in Fig.~\ref{F1-CRP}. The 
left panel gives abundance profiles for benchmark model F1 (n=10$^{3}$~cm$^{-3}$, $\chi=10$) the right
panel shows a similar model but with higher density (n=10$^{4}$~cm$^{-3}$). The higher density was 
chosen to make sure that we are outside the bistability regime. The solid lines in both panels are for
a cosmic ray helium ionization rate of $\zeta_\mathrm{CR}(\mathrm{He})=2.5\times 10^{-17}$~s$^{-1}$, the dashed lines
denote an ionization rate increased by a factor four. Different colors denote different chemical species.
The most prominent differences are highlighted with colored arrows. The factor
four in $\zeta_\mathrm{CR}(\mathrm{He})$ results in differences in density up to three orders of magnitude
in the lower density case! A detailed analysis shows that the strong abundance transitions
occur for  $\zeta_\mathrm{CR}(\mathrm{He})>8\times 10^{-17}$~s$^{-1}$. This highly
non-linear behavior disappears if we leave the critical parameter range as demonstrates in the
right panel of Fig.~\ref{F1-CRP}. \citet{boger06} emphasize that this effect is a property of the gas phase
chemical network, and is damped if gas-grain processes such as grain assisted recombination of 
the atomic ions are introduced \citep[see also][]{shalabiea95}. They conclude that
the bistability phenomenon probably does not occur in realistic dusty interstellar clouds while
\citet{lebourlot06} argues that what matters for bistability is not the number of neutralisation
channels but the degree of ionisation and that bistability may occur in interstellar clouds.
They suggest this could be one of the possible reasons of the non detection of O$_2$ by 
the ODIN satellite \citep{viti01}. Yet, another possible explanation for the absence of O$_2$ is 
freeze-out onto dust. 
However it is clear that bistability is a {\bf real} property of interstellar gas-phase 
networks and not just a numerical artifact. 
Furthermore it is important to emphasize that standard PDR models may react very sensitively
on the variation of input parameters (e.g. $\zeta_\mathrm{CR}$, the H$_2$ formation rate, the PAH 
content of the model cloud, etc.) and one has to be careful
in the interpretation of surprising model signatures.

The numerical stability and the speed of convergence may  vary significantly over different chemical networks.  Three 
major questions have to be addressed:\\
\begin{enumerate}
\item which species $i$ are to be included?
\item which reactions are to be considered?
\item which reaction rate coefficients are to be applied?
\end{enumerate}
A general answer to question 1 cannot be given, since this depends 
on the field of application.  
In steady state one has to solve a system of $M$ nonlinear equations, 
where $M$ is the number of included species, thus the
complexity of the problem scales with the number of 
species ($\propto N^2...N^3$) rather than with the number of 
chemical reactions. Nowadays CPU time
is not a major driver for the design of chemical networks. 
 Nevertheless, in some cases
a small network can give similar results as a big network. Several studies have shown 
that very large networks may include a surprisingly large number of 'unimportant' reactions, i.e. 
reactions that may be removed from the network without changing the chemical structure significantly \citep{markwick05,wakelam05a}. 
 It is more important to identify crucial species not to be omitted, i.e. species that
dominate the chemical structure under certain conditions. A well known example is the importance of sulfur for the
formation of atomic carbon at intermediate $A_\mathrm{V}$ where the charge transfer 
reaction $\mathrm{S\,+\,C^+\to\,C\,+\,S^+}$ constitutes
an additional production channel for atomic carbon, visible in a second rise in the abundance of C \citep{SD95}.
In these benchmarking calculations, sulfur was not included in order to minimize model complexity,
in spite of its importance for the PDR structure.  
 
Regarding question 2 a secure brute force approach would be the inclusion of all known reactions involving all 
chosen species, under the questionable assumption that we actually {\it know}  all important reactions
and their rate coefficients. This 
assumption is obviously invalid for grain surface reactions and gas-grain interactions
such as freeze-out and desorption.
It is important not to create artificial bottlenecks in the reaction scheme by omitting important channels.
 The choice of reaction rate coefficients depends on factors like availability, accuracy, etc.. 
A number of comprehensive databases of rate coefficients is available today, e.g. 
NSM/OHIO \citep{wakelam04,wakelam05b}, UMIST \citep{umist95,umist99}, and Meudon \citep{lebourlot93}, 
which collect the results from many different references, both theoretical and experimental.
\begin{figure}[htb]
\begin{center}
\resizebox{\hsize}{!}{\includegraphics{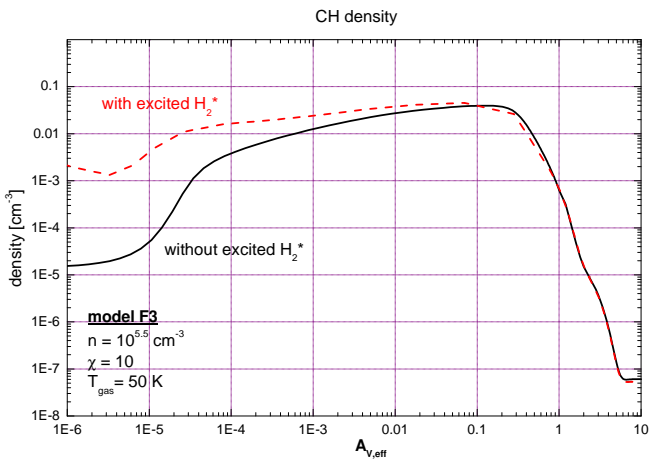}}
\caption{Comparison between model codes with (dashed line) and without (solid line)
 excited molecular hydrogen, H$_2^*$. 
The abundance profile of CH is plotted for both models against $A_\mathrm{V,eff}$. Benchmark model F3 
has a high density ($n=10^{5.5}$~cm$^{-3}$) and low FUV intensity ($\chi=10$).}\label{h2star}
\end{center}
\end{figure}

An example for the importance of explicitly agreeing on the
details of the computation of the reaction rate is the reaction:
\begin{equation}\label{chform}
\mathrm{ C\,+\,H_2\,\to\,CH\,+\,H}
\end{equation}
It has an activation energy barrier of 11700 K \citep{umist95}, effectively reducing the production of CH 
molecules. If we include vibrationally excited H$_2^*$ into the chemical network and assume 
that reaction (\ref{chform}) has no activation energy barrier for 
reactions with H$_2^*$ we obtain a significantly 
higher production rate of CH as shown in Figure \ref{h2star}.Even this 
approach is a rather crude 
assumption, but it demonstrates the importance of explicitly agreeing on how to handle the chemical calculations
in model comparisons. 

\begin{figure}[htb]
\resizebox{\hsize}{!}{\includegraphics{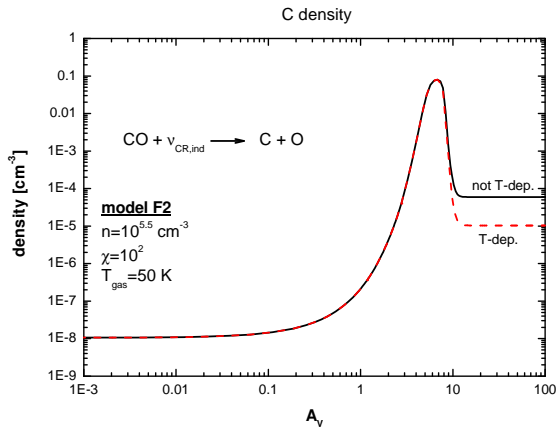}}
\caption{The density profile of atomic carbon for the benchmark model F2 (low density, high FUV, T=const=50K, as discussed in \S~\ref{benchmark_models} ). The solid curve results
from a constant dissociation by CR induced secondary photons (implicitely assuming T=300K), the 
dashed curve shows the influence of a temperature dependent
dissociation, i.e. the corresponding dissociation rate was corrected by a factor  of $\left(T/300K\right)^{1.17}$with T=50K.}\label{crphot}
\begin{center}
\end{center}
\end{figure}

Another example is the formation of C in the dark cloud part of a PDR, i.e. 
at values of $A_\mathrm{V}>\,5$. A possible formation 
channel for atomic carbon is the dissociation of CO by secondary UV 
photons, induced by cosmic 
rays \citep{umist99}. In the outer parts of the PDR the impinging FUV field dominates the dissociation of CO, 
but for high $A_\mathrm{V}$ the FUV field is effectively shielded and CR induced UV photons become 
important. For CO, this process depends on the level population of CO, and therefore is  
temperature dependent \citep{gredel87}, however this temperature dependence is
often ignored. The reaction rate  given by \citet{gredel87} has to be corrected by 
a factor of $\left(T/300K\right)^{1.17}$ effectively reducing the dissociation rate for temperatures 
below 300~K \citep{umist99}. In Figure \ref{crphot} we plot the density profile of atomic carbon for an 
isothermal benchmark model with temperature $T=50$~K. The solid 
line represents the model result for an uncorrected
photo-rate using the average reaction rate for $T=300$~K, compared to the 
results using the rate corrected for T=50 K  by $(50/300)^{1.17}$, given by the dashed curve.   

\subsubsection{Heating and Cooling}
To determine the local temperature in a cloud, the equilibrium between heating and cooling has to be 
calculated. The heating rates mainly depend on the H$_2$ formation rate, the electron density, the grain size 
distribution, grain composition, and H$_2$ treatment (i.e. two-line approximation vs. full ro-vib model), 
while the cooling rates are dominantly 
influenced by the abundance of the main cooling species and the dust 
opacity in the FIR. Table \ref{heatcool} gives an 
overview of the most important heating and cooling processes.
\begin{table}[b]
\begin{center}
\caption{Overview over the major heating and cooling processes in PDR physics}\label{heatcool}
\begin{tabular}{l|l} \hline\hline
\rule[-3mm]{0mm}{8mm}\centering{heating}&cooling\\ \hline
\rule[2mm]{0mm}{2mm}photoelectric heating (dust \& PAH) & [CII] 158$\mu$m\\
collisional de-excitation of vib. excited H$_2$&[OI] 63, 145$\mu$m\\
H$_2$ dissociation& [CI] 370, 610 $\mu$m\\
H$_2$ formation& [SiII] 35 $\mu$m\\
CR ionization  & CO,H$_2$O, OH, H$_2$\\
gas-grain collisions&Ly $\alpha$, [OI], [FeII]\\
dissipation of turbulence& gas-grain collisions \\ \hline
\end{tabular}
\end{center}
\end{table}
Most of them can be modelled at different levels of 
detail. This choice may have a major impact on the model results. One example 
is the influence of PAHs on the efficiency of the photoelectric heating, which 
results in a significantly higher temperature  e.g. at the surface of the model 
cloud if PAHs are taken into account. \citet{BT94} give convenient 
fitting formulas for the photoelectric heating. Another 
important case is the collisional de-excitation of vibrationally excited H$_2$. A detailed calculation of 
the level population shows that for temperatures above $~800$~K the lower transitions
 switch from heating to cooling. This imposes a significant influence on 
the net heating from H$_2$ vibrational de-excitation. When using an approximation for 
the heating rate it is important to account for this cooling effect \citep{roellig05}.
The cooling of the gas by line emission depends on the atomic and molecular constants as well as 
on the radiative transfer. A common approximation to the radiative 
transfer problem is by assuming escape probabilities for the cooling lines 
\citep{dejong80,stutzki84, stoerzer96}.
The excitation temperature at any point can be computed by
balancing the collisional excitation and the photon escape
probability. The local escape probability is obtained by
integrating $\exp(-\tau_\nu)$ over $4\pi$. In the escape
probability approximation it is now assumed that the radiative
interaction region is small enough so that the optical depth
in each direction is produced by molecules with the same
excitation temperature. Then the excitation problem becomes a
local one.
The [OI] 63$\mu$m line may also become very
optically thick and can act both as heating and cooling contribution. Under certain
benchmark conditions (low density, constant temperature $T_{gas}=50$~K) 
 the [OI] 63$\mu$m line even showed weak maser
behavior ( see data plots at {\small \tt http://www.ph1.uni-koeln.de/pdr-comparison}).   
Collisions between the gas particles and the dust grains also contribute to the total heating or cooling.
 
\subsubsection{Grain Properties}
Many aspects of PDR physics and chemistry are connected to dust properties.
We will give only a short overview
of the importance of dust grains in the modeling of PDRs. 
 Dust acts on the energy
balance of the ISM by means of photoelectric heating; it influences the radiative transfer by 
absorption and scattering of photons, and it acts on the chemistry of the cloud via grain 
surface reactions, e.g. the formation of molecular 
hydrogen and the depletion of other species. One distinguishes three 
dust components: PAHs, very small grains (VSGs) and 
big grains (BGs).

The properties of big grains have been reviewed recently by 
 \citet[][and references therein]{draine03}. 
 The first indirect evidence for the presence of VSGs in the ISM was presented
by \cite{andriesse78} in the case of the M17 PDR.
The dust grains themselves consist of amorphous silicates and carbonaceous material and may be
 covered with ice mantles in the denser and colder parts of the ISM. For details
 of the composition of grains and their extinction due to scattering and absorption
see \citet{li02} and references therein. 

The influence and proper treatment of 
electron densities together with grain ionization and recombination
is still to be analyzed. Not only the charge of dust and PAHs 
but also the scattering properties are still in discussion \citep{weingartner01}.  This heavily influences
the model output, e.g. the inclusion of back-scattering significantly
increases the total  H$_2$ photo-dissociation rate at the surface of the model cloud compared to
calculations with pure forward scattering.   

\subsubsection{Radiative Transfer}
The radiative transfer (RT) can be split into two distinct wavelength regimes: FUV and IR/FIR. These may also 
 be labeled as 'input' and 'output'. FUV radiation  due to
ambient UV field and/or young massive stars in the neighborhood impinges on the PDR. The 
FUV photons are absorbed on their way deeper into the cloud, giving rise to the well 
known stratified chemical structure of PDRs. In general, reemission processes can be 
neglected in the FUV, considerably simplifying the radiative transfer problem. Traveling 
in only one direction, from the edge to the inside, the local mean FUV intensity can usually be 
calculated in a few iteration steps. In contrast to the FUV, the local FIR intensity is a 
function of the temperature and level populations at all positions due to absorption and reemission 
of FIR photons. Thus a computation needs to iterate over all spatial grid points. A
common simplifying  
approximation is the spatial decoupling via the escape probability approximation. This allows to 
substitute the intensity dependence by a dependence
on the relevant optical depths, ignoring the spatial variation of the source function.
The calculation of emission line cooling then becomes primarily a problem of calculating the 
local excitation state of the particular cooling species. An overview of NLTE radiative transfer methods 
is given by \cite{vZ02}

\section{Description of the Benchmark Models}\label{benchmark_models}
\subsection{PDR Code Characteristics}
A total number of 11 model codes participated in the PDR model comparison study during 
and after the workshop in Leiden. Table \ref{participants} gives an overview 
of these codes.
\begin{table*}[thb]
\begin{center}
\caption{List of participating codes. See Appendix for short 
description of the individual models.}\label{participants}
\begin{tabular}{l|l}\hline\hline
\rule[-3mm]{0mm}{8mm}\bf{Model Name}&\bf{Authors}\\ \hline 
 {\tt Cloudy}&G. J. Ferland, P. van Hoof, N. P. Abel, G. Shaw \citep{pasp,abel05,shaw05}\\ \hline
 {\tt COSTAR}&I. Kamp, F. Bertoldi, G.-J. van Zadelhoff \citep{kamp00,kamp01}\\ \hline
 {\tt HTBKW}&D. Hollenbach, A.G.G.M. Tielens, M.G. Burton, M.J. Kaufman, M.G. Wolfire \\
&\citep{TH85,kaufman99,wolfire03} \\ \hline
 {\tt KOSMA-$\tau$}&H. St\"orzer, J. Stutzki, A. Sternberg \citep{stoerzer96}, B. K\"oster, M. Zielinsky, U. Leuenhagen\\
& \cite{bensch03},\cite{roellig05}\\ \hline
 {\tt Lee96mod}&H.-H. Lee, E. Herbst, G. Pineau des For\^ets, E. Roueff, J. Le Bourlot, O. Morata \citep{lee96}\\ \hline
 {\tt Leiden}&J. Black, E. van Dishoeck, D. Jansen and B. Jonkheid\\& \citep{black87,vDB88,jansen95}\\ \hline
 {\tt Meijerink}&R. Meijerink, M. Spaans \citep{meijerink05}\\ \hline
 {\tt Meudon}&J. Le Bourlot, E. Roueff, F. Le Petit \citep{lepetit05, lepetit02, lebourlot93}\\ \hline
 {\tt Sternberg}&A. Sternberg, A. Dalgarno \citep{SD89,SD95,boger05}\\ \hline
 {\tt UCL\_PDR}&S. Viti, W.-F. Thi, T. Bell \citep{taylor93,papadopoulos02,bell05}\\ \hline
\end{tabular}
\end{center}
\end{table*}
The codes are different in many aspects:\\
\begin{itemize}
\item finite and semi-infinite plane-parallel and spherical geometry, disk geometry
\item chemistry: steady state vs. time-dependent, different chemical reaction rates, chemical network
\item IR and FUV radiative transfer (effective or explicitly wavelength dependent), self- and mutual shielding, atomic and molecular rate coefficients
\item treatment of dust and PAHs
\item treatment of gas heating and cooling
\item range of input parameters
\item model output
\item numerical treatment, gridding, etc. 
\end{itemize}
This manifold in physical, chemical and technical differences makes it 
difficult to directly compare results from the different codes. Thus we tried 
to standardize the computation of the benchmark model clouds as much as possible. This 
required all codes to reduce their complexity and sophistication, 
often beyond what their authors considered to be acceptable, considering the actual knowledge 
of some of the physical processes.
However as the main goal of this study was to understand why and how 
these codes differ these simplifications are acceptable. Our aim was not to
provide a realistic model of real 
astronomical objects. The individual strengths (and weaknesses) of each PDR code are briefly 
summarized in the Appendix and on the website: {\small \tt http://www.ph1.uni-koeln.de/pdr-comparison} .

\subsection{Benchmark Frame and Input Values}
A total of 8 different model clouds were used for the benchmark comparison. 
The density and FUV parameter space is covered by accounting for low and high 
densities and FUV fields under isothermal conditions, giving 4 different model clouds. 
In one set of models the complexity of the model calculations was reduced
 by setting the gas and dust temperatures
to a given constant value (models F1-F4, 'F' denoting a fixed temperature), making 
the results independent of the solution
of the local energy balance. In a second benchmark set, the
thermal balance has been solved explicitly thus determining the temperature profile of the 
cloud (models V1-V4, 'V' denoting variable temperatures). Table \ref{models} gives an 
overview of the cloud parameter of all
eight benchmark clouds.
\begin{table}[thb]\begin{center}
\caption{Specification of the model clouds that were computed during the benchmark. The 
models F1-F4 use constant gas and dust temperatures, while V1-V4 have their temperatures calculated
self consistently.}\label{models}
\begin{tabular}{|c|c|} \hline
F1&F2\\
T=50~K&T=50~K\\
$n=10^3$~cm$^{-3}$, $\chi=10$&$n=10^3$~cm$^{-3}$, $\chi=10^5$\\ \hline
F3&F4\\
T=50~K&T=50~K\\
$n=10^{5.5}$~cm$^{-3}$, $\chi=10$&$n=10^{5.5}$~cm$^{-3}$, $\chi=10^5$\\ \hline
V1&V2\\
T=variable&T=variable\\
$n=10^3$~cm$^{-3}$, $\chi=10$&$n=10^3$~cm$^{-3}$, $\chi=10^5$\\ \hline
V3&V4\\
T=variable&T=variable\\
$n=10^{5.5}$~cm$^{-3}$, $\chi=10$&$n=10^{5.5}$~cm$^{-3}$, $\chi=10^5$\\ \hline
\end{tabular}
\end{center}
\end{table}

\subsubsection{Benchmark Chemistry}
One of the
crucial steps in arriving at a useful code comparison was to agree on
 the use of a standardized set of chemical species and
reactions to be accounted for. For the benchmark models we only included 
the four most abundant elements H, He, O, and C. Additionally 
only the species given in Tab.~\ref{chem1} are included in the chemical network calculations:\\

\begin{table}[tbph]\begin{center}
\caption{Chemical content of the benchmark calculations.}\label{chem1}
\begin{tabular}{l} 
\hline \hline
\rule[-3mm]{0mm}{8mm}\bf Chemical species in the models\\ \hline
\rule[2mm]{0mm}{2mm}H, H$^+$, H$_2$, H$_2^+$, H$_3^+$\\
 O, O$^+$, OH$^+$, OH, O$_2$, O$_2^+$, H$_2$O, H$_2$O$^+$, H$_3$O$^+$\\
 C, C$^+$, CH, CH$^+$, CH$_2$, CH$_2^+$, CH$_3$,\\
 CH$_3^+$, CH$_4$, CH$_4^+$, CH$_5^+$, CO, CO$^+$,HCO$^+$\\
He, He$^+$, e$^-$\\ \hline
\end{tabular}

\end{center}
\end{table}

The chemical reaction rates 
are taken from the UMIST99 database \citep{umist99} together with some corrections 
suggested by A. Sternberg. The complete 
reaction rate file is available online ({\tt \small http://www.ph1.uni-koeln.de/pdr-comparison}).
 To reduce the overall modeling complexity, PAHs were neglected in the
chemical network and were only
considered for the photoelectric heating \citep[photoelectric heating 
efficiency as given by][]{BT94} in models V1-V4. Codes
which calculate time-dependent chemistry used a suitably long time-scale
in order to reach steady state (e.g. UCL\_PDR used 100 Myr). 

\subsubsection{Benchmark Geometry}
All model clouds are plane-parallel, semi-infinite clouds of constant total hydrogen
 density  $n = n(\mathrm{H})+2\,n(\mathrm{H}_2)$. Spherical codes approximated this by assuming a very
large radius for the cloud.   

\subsubsection{Physical Specifications}
As many model parameters as possible were agreed upon at the start of the 
benchmark calculations, to avoid 
confusion in comparing model results. To this end we set the most crucial
model parameters to the following values:
the value for the standard UV field was taken as $\chi=10$ and $10^5$ times the Draine (1978) field.
For a semi-infinite
plane parallel cloud the CO dissociation rate at the
cloud surface for $\chi=10$ should equal $10^{-9}$~s$^{-1}$, using that for
optically thin conditions (for which a point is exposed
to the full $4 \pi$ steradians, as opposed to  $2 \pi$
at the cloud surface) the CO dissociation rate is
$2\times 10^{-10}$~s$^{-1}$ in a unit Draine field. The cosmic ray H ionization rate is assumed to 
be $ \zeta = 5\times 10^{-17}$~s$^{-1}$ and the 
visual extinction  $A_\mathrm{V} = 6.289\times 10^{-22} N_{\mathrm{H,tot}}$.
If the codes do not
explicitly calculate the H$_2$ photo-dissociation
rates (by summing over oscillator strengths etc.) we
assume that the unattenuated H$_2$ photo-dissociation rate
in a unit Draine field is equal to $5.18\times 10^{-11}$~s$^{-1}$,
so that at the surface of a semi-infinite cloud
for 10 times the Draine field the H$_2$ dissociation rate is $2.59\times 10^{-10}$~s$^{-1}$ (numerical values from {\tt Sternberg}. See \S~\ref{F1-F4}
for a discussion on H$_2$ dissociation rates).
For the dust attenuation factor in the H$_2$ dissociation
rate we assumed $\exp(-k\,A_\mathrm{V})$ if not treated explicitly wavelength dependent. The value $k=3.02$ 
is representative for the effective opacity in the 912-1120 $\AA$ range (for a specific value of
$R_\mathrm{V}\approx 3$). We use 
a very simple H$_2$ formation rate coefficient
$R = 3\times 10^{-18}\,T^{1/2} = 2.121\times^{-17}$~cm$^3$~s$^{-1}$ \citep{SD95} at $T=50$~K, assuming
that every atom that hits a grain sticks and reacts to H$_2$. A summary of 
the most important model parameters is given in Table \ref{phys1}.

\begin{table}[b]
\begin{center}
\caption{Overview of the most important model parameter. All abundances
are given w.r.t. total H abundance.}\label{phys1}
\begin{tabular}{lll}
\hline\hline
\multicolumn{3}{c}{\rule[-3mm]{0mm}{8mm}\bf Model Parameters}\\ \hline
\rule[2mm]{0mm}{2mm}A$_\mathrm{He}$&0.1&elemental He abundance\\
A$_\mathrm{O}$&$3\times 10^{-4}$&elemental O abundance\\
A$_\mathrm{C}$&$1\times 10^{-4}$&elemental C abundance\\
$\zeta_{CR}$&$5\times 10^{-17}$~s$^{-1}$&CR ionization rate\\
$A_\mathrm{V}$&$6.289\times 10^{-22} N_\mathrm{H_{total}}$&visual extinction\\
$\tau_\mathrm{UV}$&$3.02 A_v$&FUV dust attenuation\\
$v_b$ & 1~km~s$^{-1}$&Doppler width\\
$D_\mathrm{H_2}$&$5\times 10^{-18}\cdot \frac{\chi}{10}$~s$^{-1}$&H$_2$ dissociation rate\\
$R$&$3\times 10^{-18} T^{1/2}$~cm$^{3}$~s$^{-1}$&H$_2$ formation rate\\
$T_\mathrm{gas,fix}$&50~K&gas temperature (for F1-F4)\\
$T_\mathrm{dust,fix}$&20~K&dust temperature (for F1-F4)\\
$n$&$10^3, 10^{5.5}$~cm$^{-3}$&total density\\
$\chi$&$10,10^5$&FUV intensity w.r.t.\\& & \cite{draine78} field\\
& & (i.e.$\chi=1.71~G_0$)\\
\hline\\
\end{tabular}
\end{center}
\end{table}

\section{Results} \label{results} 
\begin{figure*}
\centering
\includegraphics[width=16 cm]{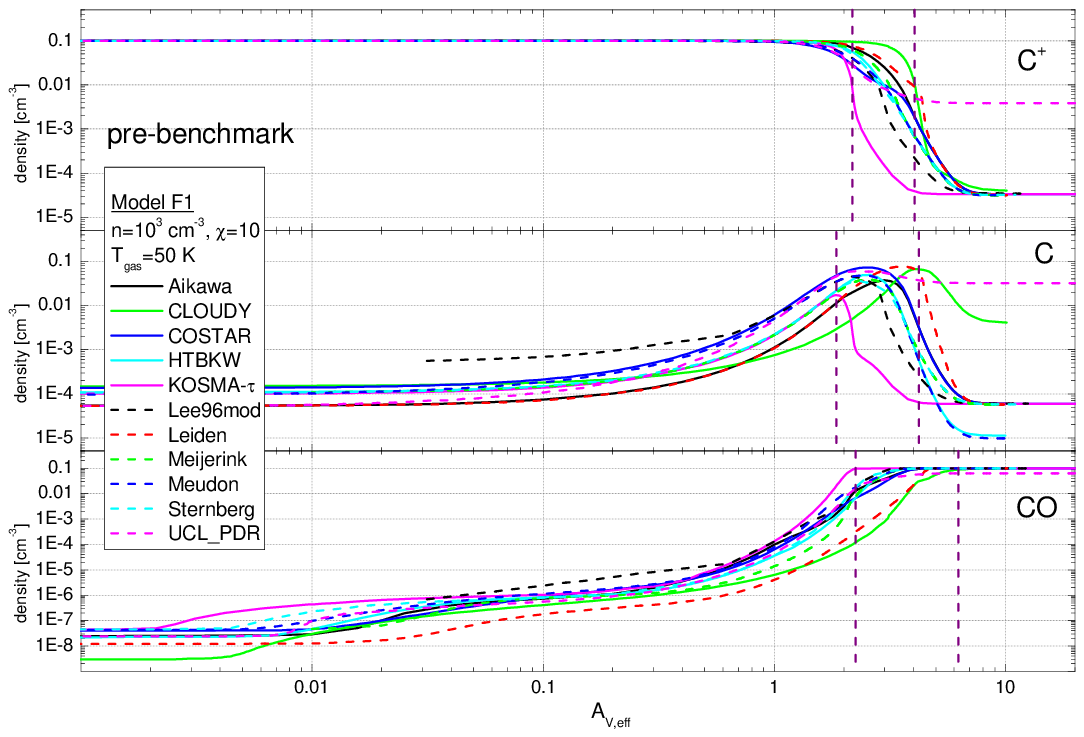}
\includegraphics[width=16 cm]{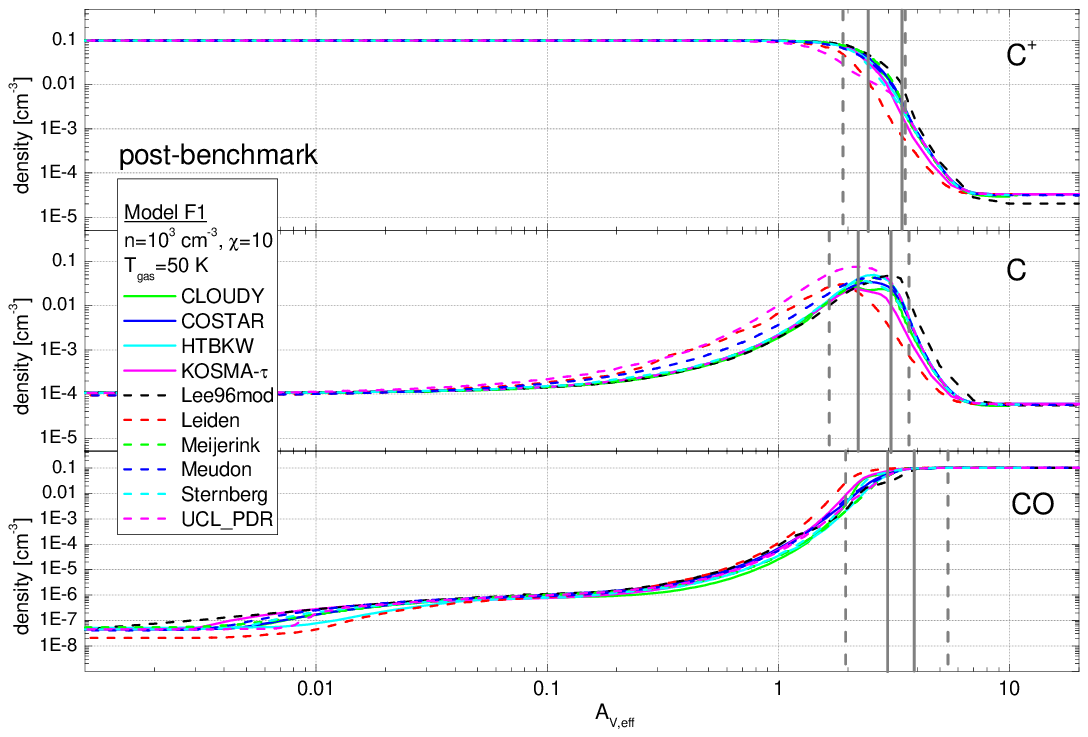}
\caption{Model F1 (n=10$^{3}$~cm$^{-3}$, $\chi=10$): Comparison between the density profiles of C$^+$ (top), C (middle), and
CO (bottom) before (top) and after (bottom) the comparison study. The vertical lines 
indicate the code dependent scatter. For C and CO they 
indicate the depths at which the maximum density is reached, while for C$^+$ they indicate the depths at which the density 
dropped by a factor of 10. Dashed lines indicate pre-benchmark results, while solid lines are post-benchmark.}\label{F1-C-both}
\end{figure*}
In the following section we give a short overview of the up to date results of the
PDR model comparison. The names of the model codes are printed in typewriter font
(e.g. {\tt COSTAR}). We will refer to the two stages of the benchmarking
results by pre- and post-benchmark, denoting the model results at the beginning 
of the comparison and at its end respectively. All pre- and post-benchmark
results are posted at {\small \tt http://www.ph1.uni-koeln.de/pdr-comparison}. One model from the 
initial 12 participating model was left out in the post-benchmark plots because the authors could 
not attend the workshop. 
In addition, the KOSMA-$\tau$ models \citep{roellig05} and the models
by Bensch, which participated in the comparison as seperate codes, have 
been merged to a single set (labeled KOSMA-$\tau$) as
they are variants on of the same basic model which do not differ
for the given benchmarking parameter set, and consequently
give identical results. To demonstrate
the impact of the benchmark effort on the results of the participating PDR codes we plot the 
well known C/~C$^+\,$/~CO transition for a typical PDR environment before and after the
changes identified as necessary during the
benchmark in Fig.~\ref{F1-C-both}. The photo-dissociation of carbon monoxide 
is thought to be well understood for almost 20 years \citep{vDB88}. Nevertheless we see a significant
scatter for the densities of C$^+$, C, and CO in the top plot of Fig.~\ref{F1-C-both}.
The scatter in the pre-benchmark rates is significant. Most deviations 
could be assigned to either bugs in the pre-benchmark codes, misunderstandings, or to incorrect 
geometrical factors (e.g. $2\,\pi$ vs. $4\,\pi$).
This emphasizes the importance of this comparative study to establish a  
uniform understanding about how to calculate even these basic figures. 
Despite the considerable current interest because of, e.g. SPITZER
results, we do not give the post-benchmark predictions for the H$_2$
mid-IR and near IR lines (or the corresponding Boltzmann diagram).
Only a small fraction of the participating codes is able to compute the detailed
H$_2$ population and emission, and the focus of this analysis is the comparison
between the benchmark codes.

\subsection{Models with Constant Temperature F1-F4}\label{F1-F4}
The benchmark models F1 to F4 were calculated for a fixed gas temperature
of 50~K. Thus, neglecting any numerical issues, all differences in the chemical structure 
of the cloud are due to the different photo-rates, or non-standard chemistry. Some PDR codes
used slightly different chemical networks. The code {\tt Sternberg} uses the standard chemistry with the
addition of vibrational excited hydrogen and a smaller H-H$_2$ formation network. 
The results by  {\tt Cloudy} were obtained with two 
different chemical setups: The pre-benchmark chemistry had the chemical network of \citet{TH85}.  The 
post benchmark results use the corrected UMIST database. {\tt Cloudy} also used a different set of radiative 
recombination coefficients for the pre-benchmark calculations which were the major source
for their different results \citep{abel05}. The carbon photoionization and radiative recombination rates
are very sensitive to radiative transfer and hence to dust properties. The
dust properties in {\tt Cloudy} are different from what is implicitly assumed
in the UMIST fits. {\tt Cloudy's} post-benchmark results are achieved after switching to 
the benchmark specifications. After the switch they agree very well with the
other codes.
\begin{figure*}
\centering
\includegraphics[width=17 cm]{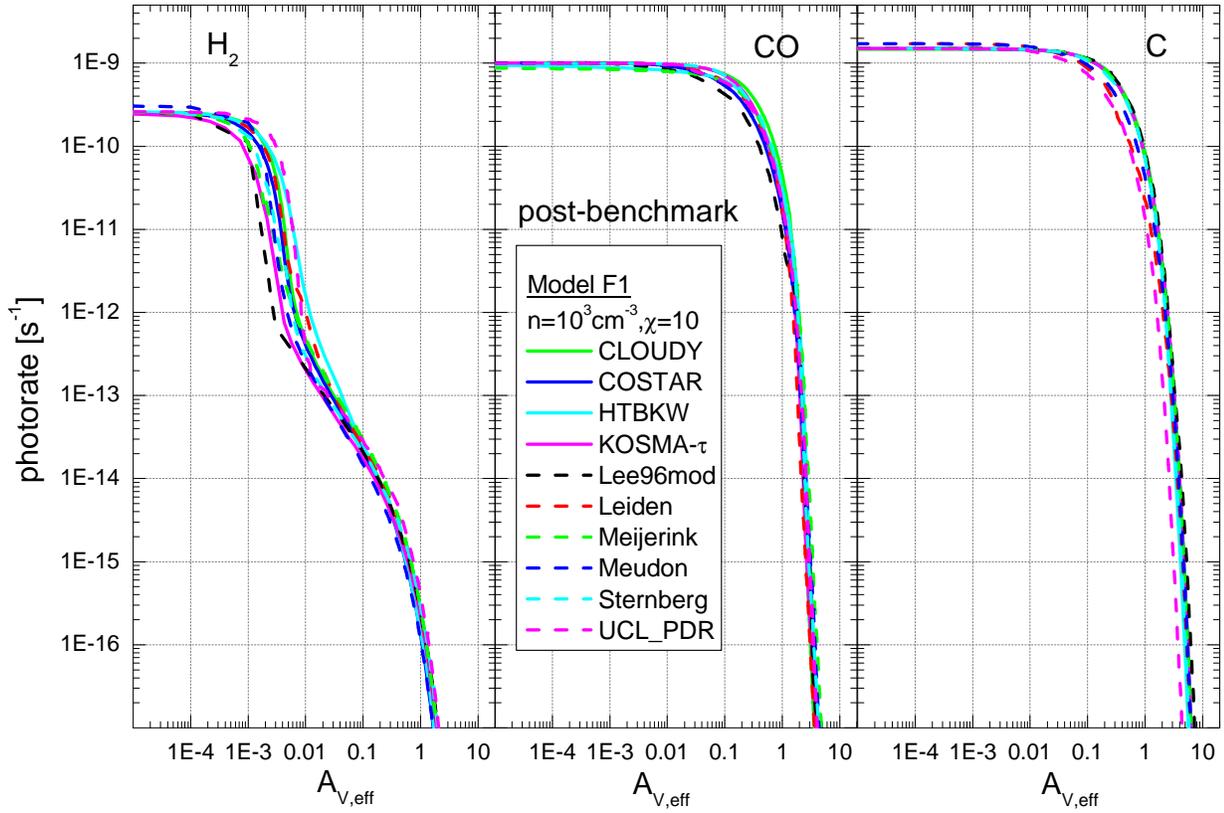}
\caption{Model F1 (n=10$^{3}$~cm$^{-3}$, $\chi=10$):  The photo-dissociation rates of
H$_2$ ( left column), of CO (middle column) and the photo-ionization rate of C ( right column) 
after the comparison study.}\label{F1-photorates-both}
\end{figure*}
\begin{figure}
\centering
\resizebox{\hsize}{!}{\includegraphics{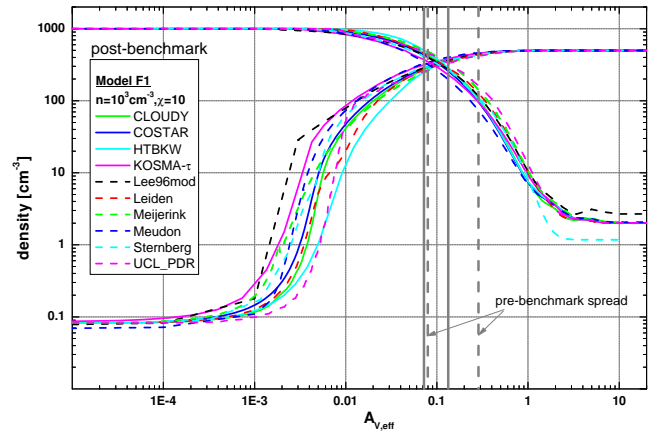}}
\caption{Model F1 (n=10$^{3}$~cm$^{-3}$, $\chi=10$) The H-H$_2$ transition zone
after the comparison study. Plotted is the number density of atomic and molecular hydrogen as a function of 
A$_\mathrm{V,eff}$. The vertical lines denote the 
range of the predicted transition depths for pre- and post-benchmark results (dashed and solid lines respectively).}\label{F1-H-both}
\end{figure}
In Fig.~\ref{F1-C-both} we present the pre- and post-benchmark results for the main carbon bearing
 species C$^+$, C, and CO. To emphasize the pre-to-post changes we added several vertical marker 
lines to the plots. For C and CO they 
indicate the depths at which the maximum density is reached, while for 
C$^+$ they indicate the depths at which the density 
has dropped by a factor of 10. Dashed lines indicate pre-benchmark results, 
while solid lines are post-benchmark. 
In the pre-benchmark results the code dependent scatter for these depths is
 $\Delta\,A_\mathrm{V,eff}\approx 2-4$ and drops to $\Delta\,A_\mathrm{V,eff}\approx 1$ 
in the post-benchmark results. 

In the post-benchmark results, the  {\tt Leiden} and {\tt UCL\_PDR} models show a slightly different behavior.
The predicted peak depth of C  is
somewhat smaller than for the other codes. The peak C density of {\tt UCL\_PDR} is  roughly 
50\% higher than in the other codes.
A comparison with the photo-ionization of C shown in 
Fig.~\ref{F1-photorates-both} confirms that a slightly stronger shielding for the 
ionization of C is the reason for the different
behavior of C and C$^+$.  The dark cloud 
densities for C$^+$, C, and CO agree very well, except for a somewhat 
smaller C$^+$ density in the  {\tt Lee96mod} results.

In Fig.~\ref{F1-photorates-both} we plot the post-benchmark photo-rates 
for dissociation of H$_2$ (left column)  and CO (middle column)
and for the ionization of C (right column), computed for model F1. Even 
for this simple model there are some significant differences
between the models in the various rates. In the pre-benchmark results, several codes 
calculated different photo-rates at the edge of the
model cloud, i.e. for very low values of A$_\mathrm{V,eff}$.
Some codes calculated surface
photo-dissociation rates between $4-5\times 10^{-10}$~s$^{-1}$ compared to the expected value of
 $2.59\times 10^{-10}$~s$^{-1}$. Most of these deviations were 
due to exposure to the full $4\pi$ steradians FUV field instead the correct $2\pi$, but also due to 
different effects, like the FUV photon back-scattering in the {\tt Meudon} results.
The pre-benchmark rates of  {\tt KOSMA-$\tau$} were shifted toward slightly 
lower values of A$_\mathrm{V}$ because of an incorrect scaling 
between A$_\mathrm{V}$ and A$_\mathrm{V,eff}$ and an incorrect 
calculation of the angular averaged photo-rate (the model features a spherical geometry
with isotropic FUV illumination).
 The  post-benchmark results (Fig.~\ref{F1-photorates-both}) show
 that most deviations have been corrected. The remaining offset
for the  {\tt Meudon} result is due to the consideration of backscattered FUV photons, increasing the
local mean FUV intensity.  The pre- to post-benchmark changes for the photo-rates of CO and C are
even more convincing (see online archive). The post-benchmark results are in 
very good agreement  except for some minor difference, e.g.  {\tt UCL\_PDR's}
 photo-ionization rate of C showing some deviation from the main field.
 
The depth-dependence of the H$_2$ photo-dissociation rate is reflected in the 
structure of the H-H$_2$ transition zone. Fig.~\ref{F1-H-both}
shows the densities of atomic and molecular hydrogen after the benchmark. The vertical lines 
denote the minimum and maximum transition depths before (dashed) and after the benchmark (solid). In 
the pre-benchmark results
the predicted transition depth ranges from 0.08 A$_\mathrm{V,eff}$ to 0.29 
A$_\mathrm{V,eff}$. In the post-benchmark results
this scatter is reduced by more than a factor of 3.
  {\tt Sternberg} gives a slightly smaller H density in the dark cloud part. 
In this code, cosmic ray (CR) destruction 
and grain surface formation are the only reactions considered in the calculation of the H$_2$ density.
The other codes use additional reactions.  The reactions:
\begin{eqnarray}
& &\mathrm{H_2^+\,+\,H_2\,\to\,H_3^+\,+\,H}\;\;\;(k=2.08\times 10^{-9}\,\mathrm{cm^{3}\,s^{-1}})\nonumber\\
& &\mathrm{H_2\,+\,CH_2^+\,\to\,CH_3^+\,+\,H}\;\;\;(k=1.6\times 10^{-9}\,\mathrm{cm^{3}\,s^{-1}})\nonumber
\end{eqnarray}
contribute to the total H density at high $A_\mathrm{V,eff}$. 
This results in a somewhat higher H density as shown
 in Fig.~\ref{F1-H-both}. The  {\tt Meudon} model  gives a slightly smaller
 H$_2$ density at the edge of the cloud than the other 
codes. This is due to the already mentioned
higher photo-dissociation rate of molecular hydrogen applied in their calculations. 

\begin{figure*}
\centering
\includegraphics[width=16 cm]{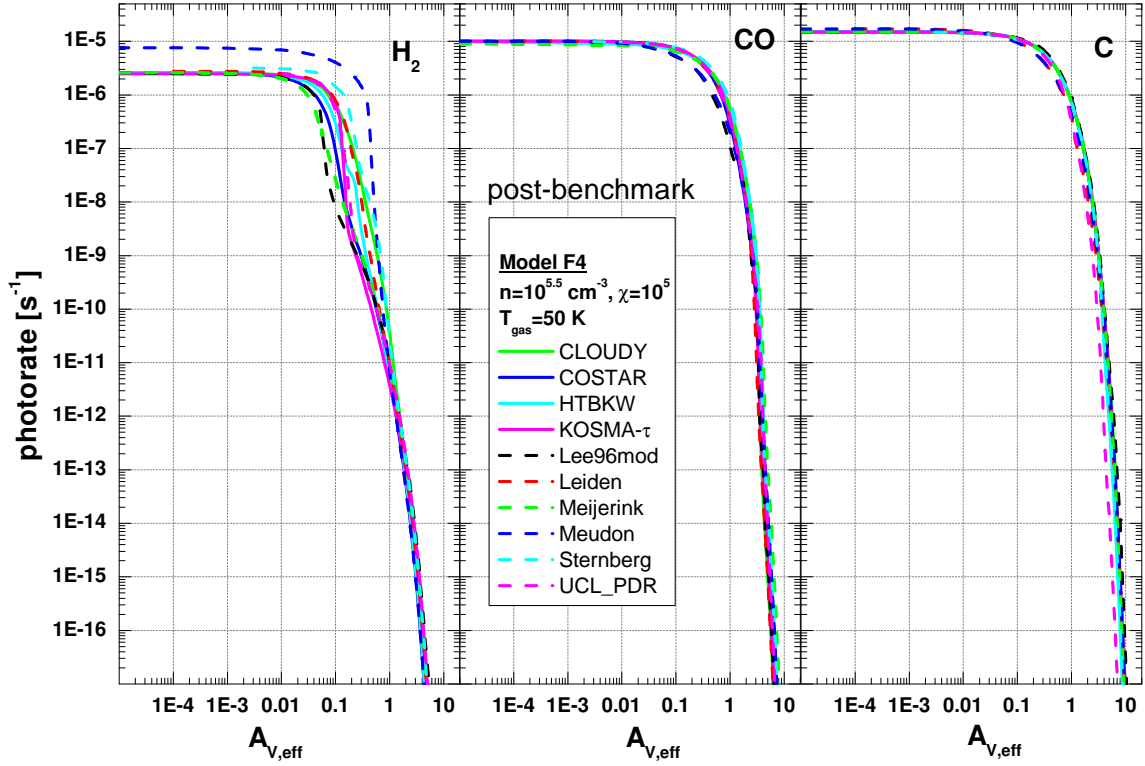}
\caption{Model F4 (n=10$^{5.5}$~cm$^{-3}$, $\chi=10^5$): The post-benchmark photo-dissociation rates of
H$_2$ (left column), of CO (middle column) and the photo-ionization rate of C (right column) 
(upper plot).}\label{F4-photo}
\end{figure*}

The model F1 may represent a typical translucent cloud PDR, e.g., the line of sight
toward HD 147889 in Ophiuchus \citep{liseau99}. The low density and FUV intensity conditions
emphasize some effects that would be hard to notice otherwise. This includes purely numerical issues like
gridding and interpolation/extrapolation of shielding rates. These differences explain why the various codes
still show some post-benchmark scatter. We  relate differences in the predicted abundances
to the corresponding rates for ionization and dissociation.

 Since most of the codes use the same
chemical network and apply the same temperature, the major source for remaining deviations should be
related to the FUV radiative transfer. To study this we present some results of benchmark model
F4 featuring a density  $n=10^{5.5}$~cm$^{-3}$ and a FUV intensity $\chi=10^5$, in order to enhance
any RT related differences and discuss them in more detail.   
\begin{figure*}
\centering
\includegraphics[width=16 cm]{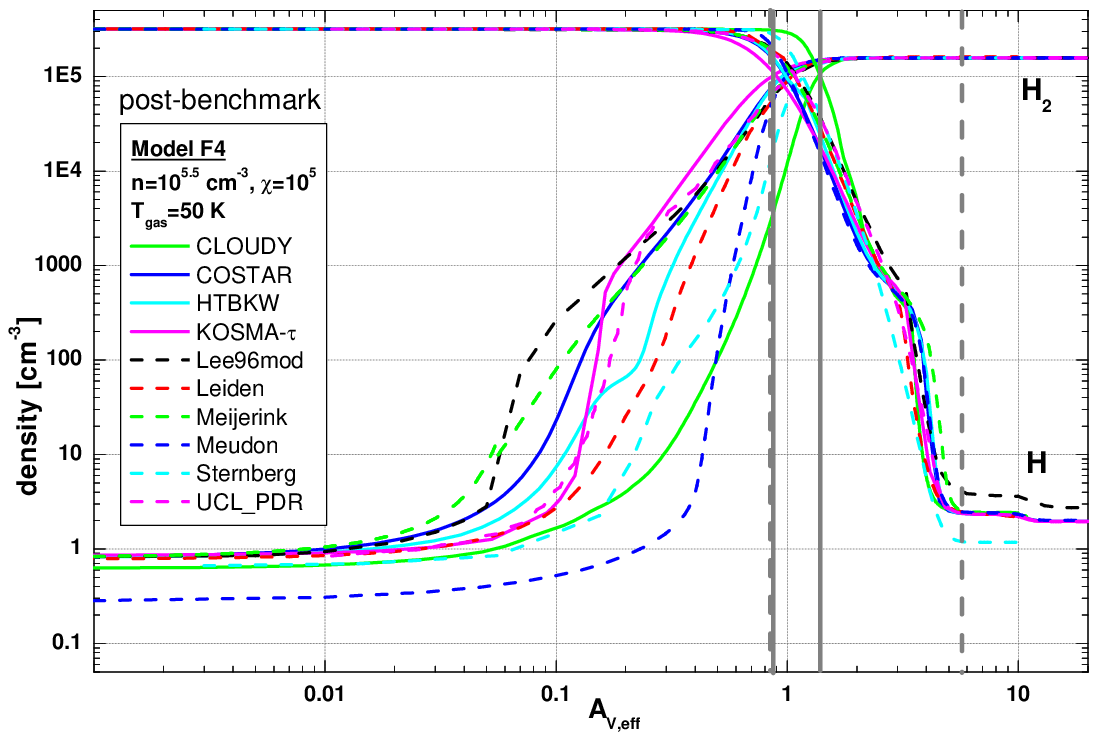}
\includegraphics[width=16 cm]{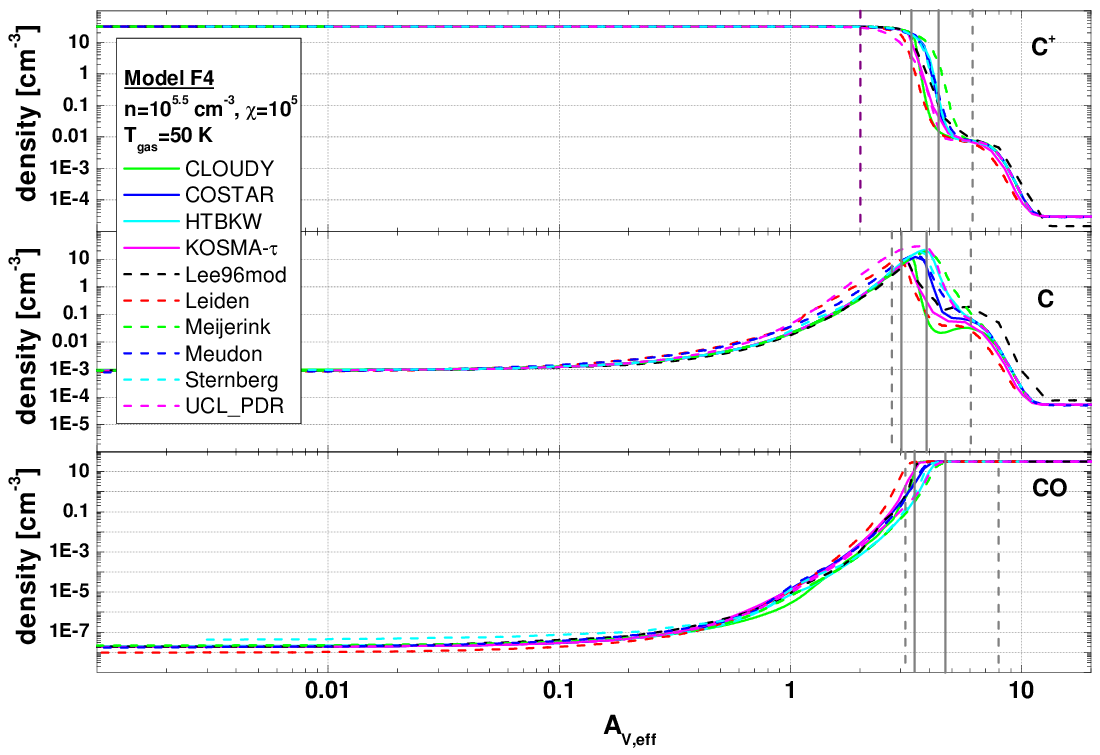}
\caption{Model F4 (n=10$^{5.5}$~cm$^{-3}$, $\chi=10^5$): The upper panel shows the post-benchmark results for the H and
H$_2$ densities. The lower panel shows the post-benchmark density profiles of C$^+$, C, and CO. 
 The vertical gray lines in both panels indicate the pre-to-post changes.}\label{F4-H-C-both}
\end{figure*} 
Fig.~\ref{F4-photo} shows the post-benchmark photo-rates for the model F4. 
The higher unshielded H$_2$ photo-rate in the
 {\tt Meudon} results, already visible in model F1 (Fig.~\ref{F1-photorates-both}) is now significantly 
enhanced due to the increased FUV flux. 
 {\tt Meudon}, as well as {\tt Cloudy}, {\tt Leiden} and {\tt Sternberg}, treat the 
 hydrogen molecule by calculating the local level population
and determining the photo-dissociation rate by integrating each absorption line over the 
absorption cross section and summing over all absorption lines. {\tt Meudon}, {\tt Cloudy}, and {\tt Leiden}
integrate the line profile over the attenuated
spectrum, in order to account for line overlap effects, while {\tt Sternberg} treats each line seperately,
neglecting line overlap.
Most other codes just assume that the photodissociation scales with the incident radiation field, neglecting
any influence from varying H$_2$ level populations. 
 One reason for the different H$_2$ photo-rate is a different local mean FUV intensity, caused
by backscattered photons. However, this should only account for approximately 10\% of the 
increased dissociation rate.
The remaining differences are due to different treatment of H$_2$. Either they use  
different equations,  
e.g. full ro-vib resolution in {\tt Meudon} and {\tt Sternberg} vs. only vib. population in {\tt KOSMA-$\tau$}, or they apply
different molecular data. {\tt Sternberg} uses data from \citet{SD89,SN99}. {\tt Meudon} 
uses collisional data from \citet{flower97,flower98,flower99} and associated papers, and radiative data
from \citet{abgrall00}, including dissociation efficiencies. These different data sets result in:
\begin{enumerate}
\item Excited rotational states are much more populated in {\tt Meudon's} 
results than in {\tt Sternberg}
\item Dissociation from an excited rotational level increases much faster with
J in {\tt Meudon's} data.
\end{enumerate}
Both effects lead to dissociation probabilities that differ by 2-3 in case of Model F4.
Due to the structure of the code these features could not be turned off in
{\tt Meudon} results.

 The photo-rates for CO and C are in very good accord, but we notice a
considerable spread in the shielding behavior of the hydrogen photo-rate. This spread is due to
the particular implementation of H$_2$ shielding native to every code, by either using tabulated shielding 
functions or explicitly calculating the total cross section at each wavelength. 
The different photo-rates directly cause a
different H-H$_2$ transition profile, shown in the top panel of Fig.~\ref{F4-H-C-both}. 
The low molecular hydrogen 
densities in the {\tt Meudon} and  {\tt Cloudy} models are again due to the higher H$_2$ photo-dissociation rate.
{\tt Sternberg's} slightly lower H$_2$ abundance at the edge of the cloud is
 consistent with the marginally higher,
unshielded H$_2$ photo-dissociation rate, seen in the top 
plot in Fig.~\ref{F4-photo}.  The  {\tt Meijerink} code shows 
the earliest drop in the photo-rate, reflected by the corresponding increase in the H$_2$ density.  The 
qualitatively different H$_2$ profile in {\tt KOSMA-$\tau$} is most likely due
to the spherical geometry in the code.  
 Again  {\tt Sternberg} produces slightly smaller H densities for high values of $A_\mathrm{V,eff}$. Since 
 {\tt Sternberg} does not consider additional reactions for the H/H$_2$ balance its H density profile is 
the only one not showing the slight kink at $A_\mathrm{V,eff}\approx 2...3$.  These
deviations do not significantly change the total column density of hydrogen. Hence the impact on 
any comparison with observational findings is small. Nevertheless one would expect that under
the standardized benchmark conditions all codes produce very similar results, yet we note a considerable spread
in hydrogen abundances for $A_\mathrm{V,eff}>2$. This again emphasizes how 
complex and difficult the numerical modeling of PDRs is. The bottom panel in Fig.~\ref{F4-H-C-both} shows
the density profiles of C$^+$, C, and CO. Here, the different codes are in good agreement. The largest spread
is visible for the C density between $A_\mathrm{V,eff}\approx 3...6$. The results for C$^+$ and CO differ less. 
{\tt Lee96mod's} results for C$^+$ and C show a small offset for $A_\mathrm{V,eff} > 6$. They produce
slightly higher C abundances and lower C$^+$ abundances in the dark cloud part. The different codes agree
very well in the predicted depth where most carbon is locked up in CO 
($A_\mathrm{V,eff}\approx 3.5...4.5$). This range
improved considerably compared to the pre-benchmark predictions of $A_\mathrm{V,eff}\approx 3...8$.  
    
The results from models F1-F4 clearly demonstrate the importance of the  PDR code
 benchmarking effort. The pre-benchmark
results show a significant code-dependent scatter. Although many of these deviations have been
removed during the benchmark activity, we did not achieve identical results with different codes. Many 
uncertainties remained 
even in the isothermal case, raising the need for a deeper follow up study.
\subsection{Models with Variable Temperature V1-V4}
\begin{figure}
\centering
\resizebox{\hsize}{!}{\includegraphics{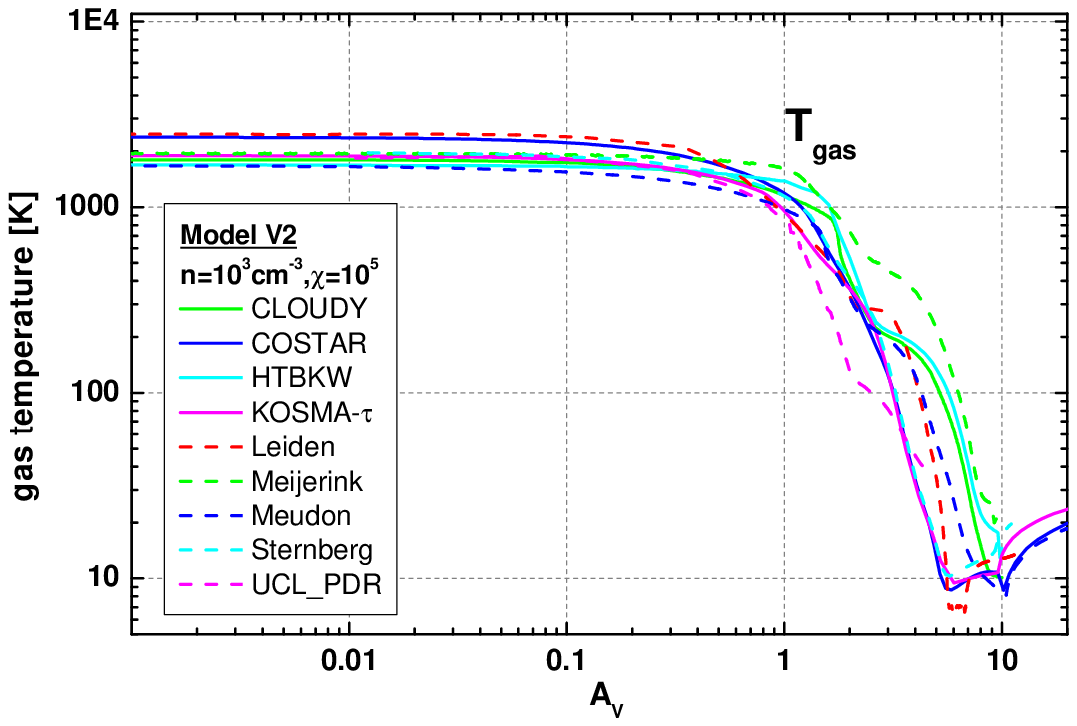}}
\caption{Model V2 (n=10$^{3}$~cm$^{-3}$, $\chi=10^5$): The plot shows 
the post-benchmark results for the gas temperature.}\label{V2-temp}
\end{figure}

\begin{figure*}
 \centering
 \includegraphics[width=16 cm]{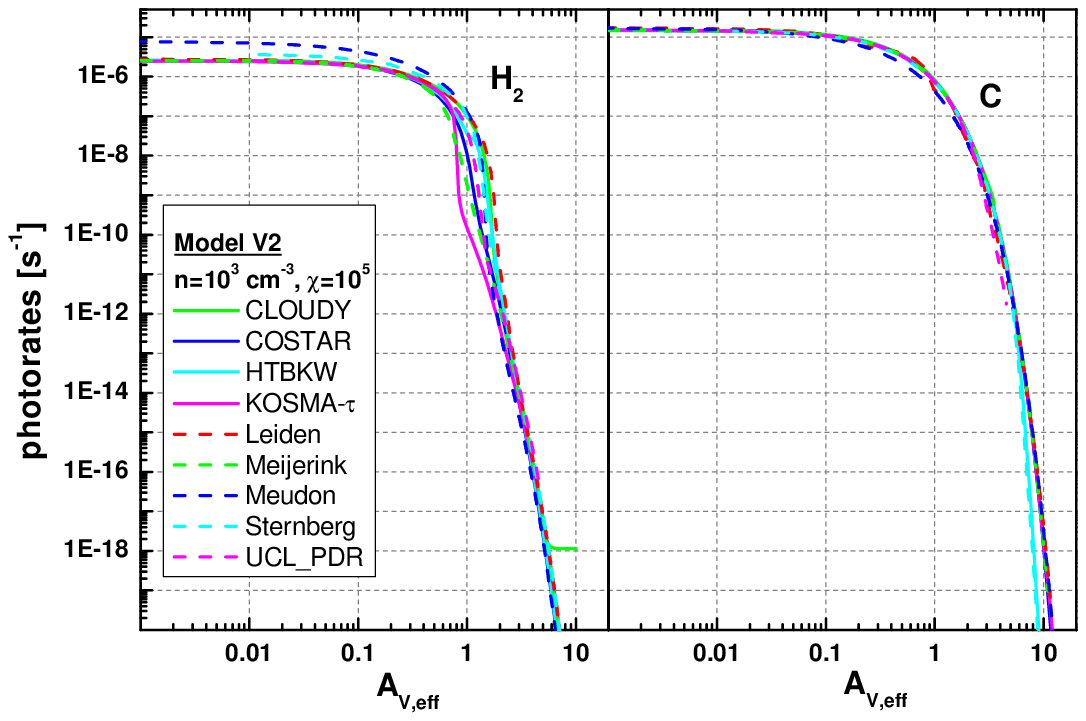}
 \includegraphics[width=16 cm]{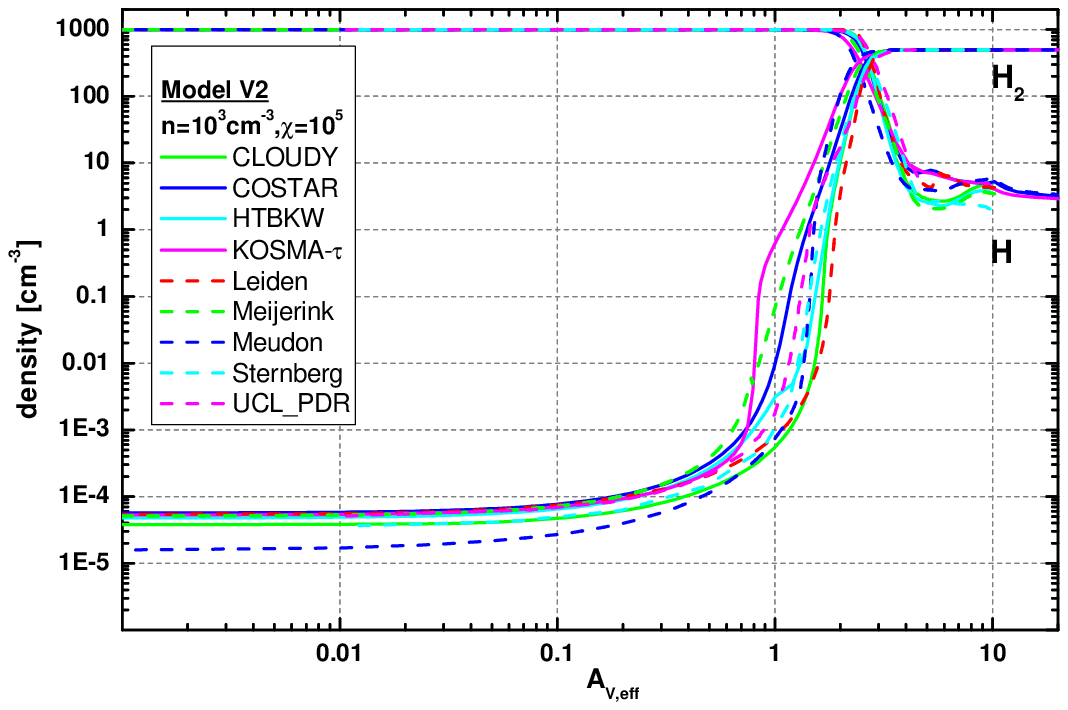}
 \caption{Model V2 (n=10$^{3}$~cm$^{-3}$, $\chi=10^5$): The post-benchmark photo-dissociation rates of
 H$_2$ (left column), and the photo-ionization rate of C (right column)
 (upper plot). The lower plots shows the  H and
 H$_2$ densities.}\label{V2-photo-H-both}
 \end{figure*}

\begin{figure*}
 \begin{minipage}[hbt]{8.5cm}
 \centering
 \includegraphics[width=8.5 cm]{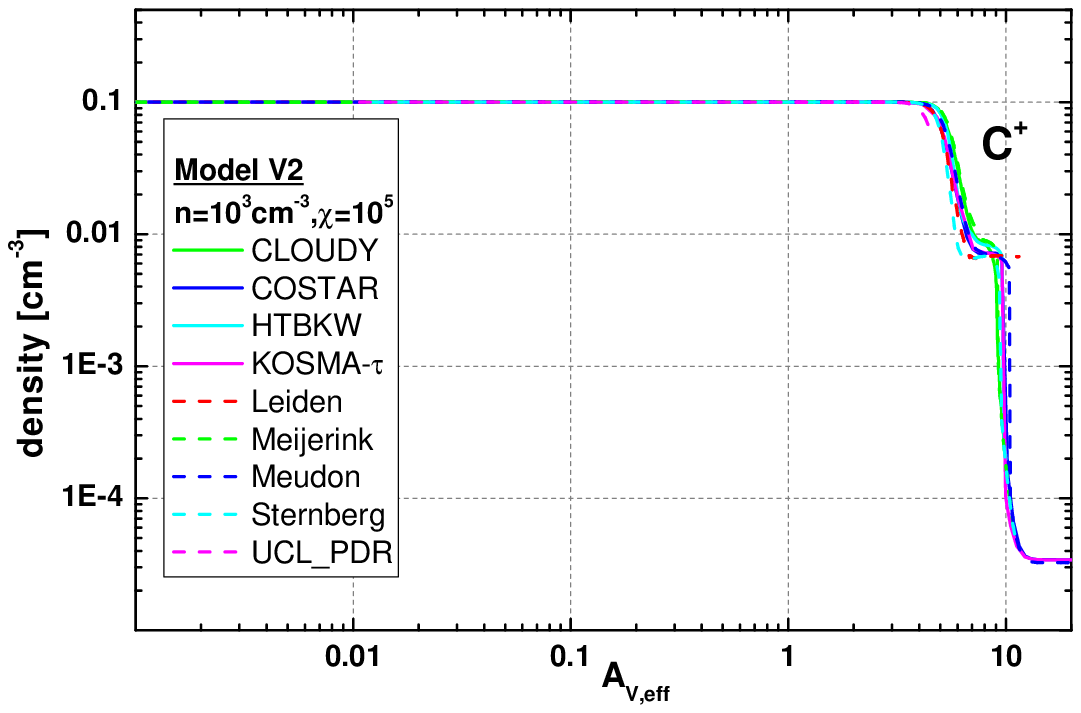}
 \end{minipage}
 \hfill
 \begin{minipage}[hbt]{8.5cm}
 \centering
 \includegraphics[width=8.5 cm]{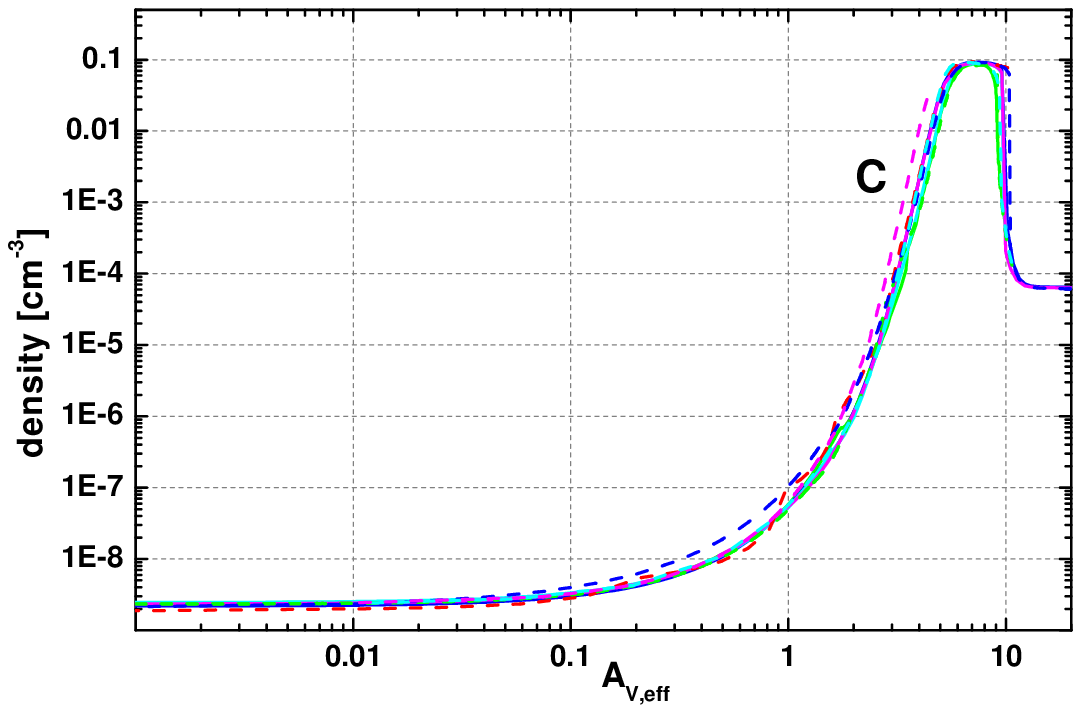}
 \end{minipage}
 \begin{minipage}[hbt]{8.5cm}
 \centering
 \includegraphics[width=8.5 cm]{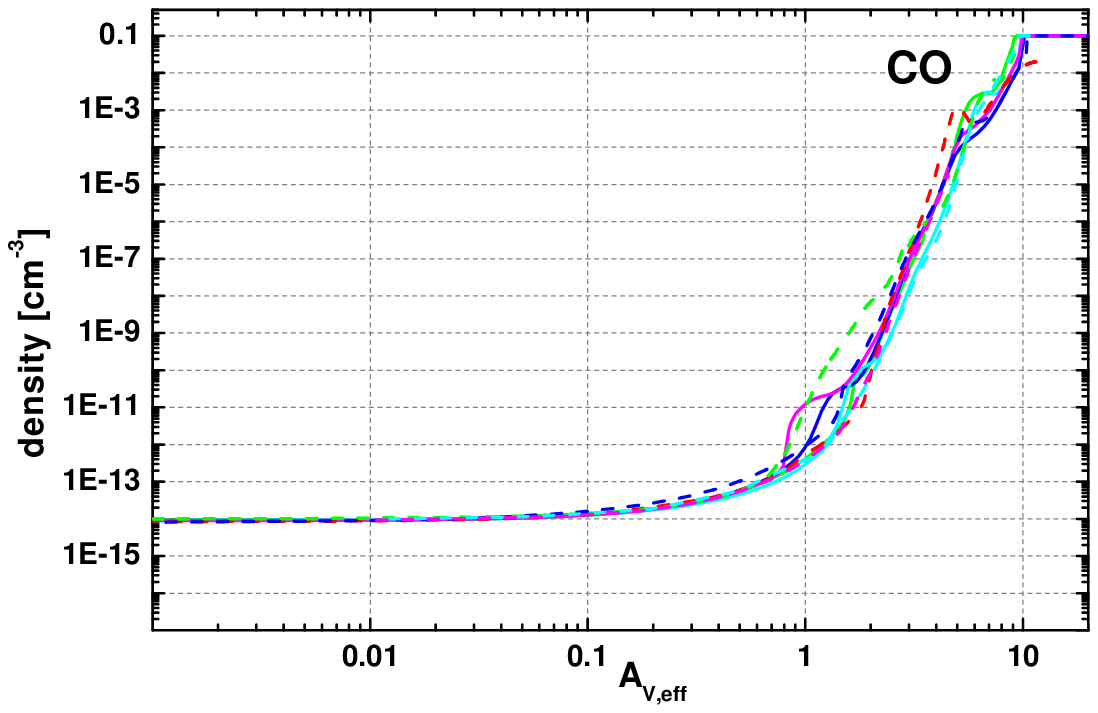}
 \end{minipage}
 \hfill
 \begin{minipage}[hbt]{8.5cm}
 \centering
 \includegraphics[width=8.5 cm]{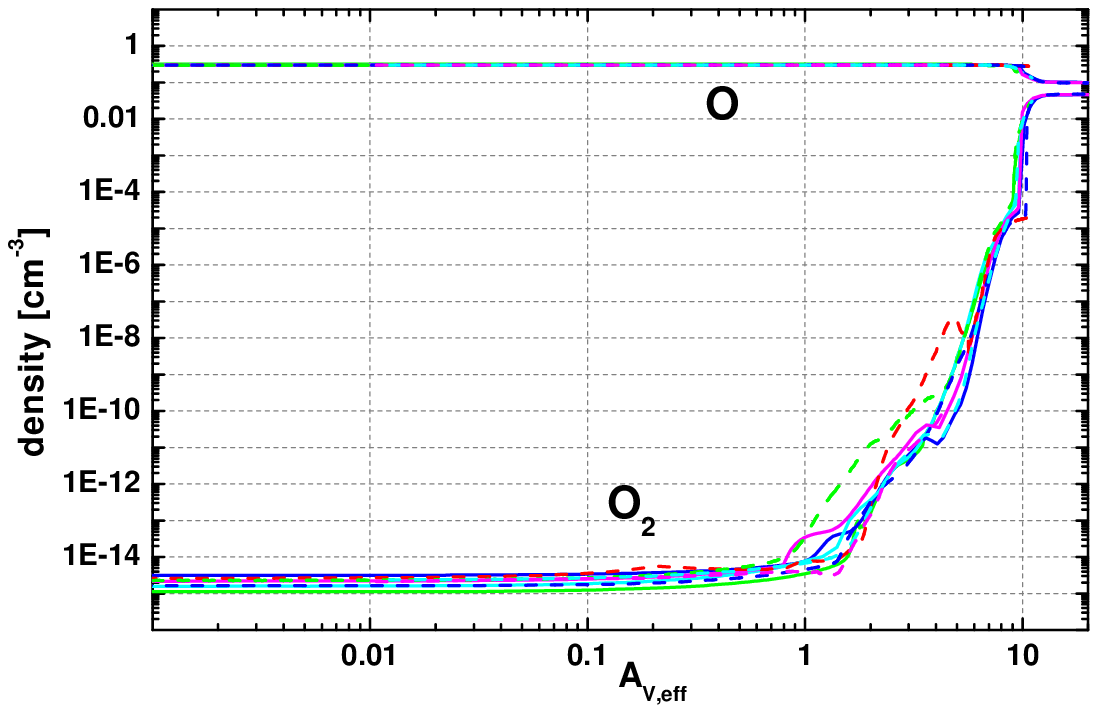}
 \end{minipage}
 \caption{Model V2 (n=10$^{3}$~cm$^{-3}$, $\chi=10^5$): The post-benchmark results 
for the densities of C$^+$ (top left),
 the densities of C (top right), and the densities of CO (bottom left) and O and O$_2$ 
(bottom right).}\label{V2-big}
 \end{figure*}

\begin{figure}
 \centering
\resizebox{\hsize}{!}{\includegraphics{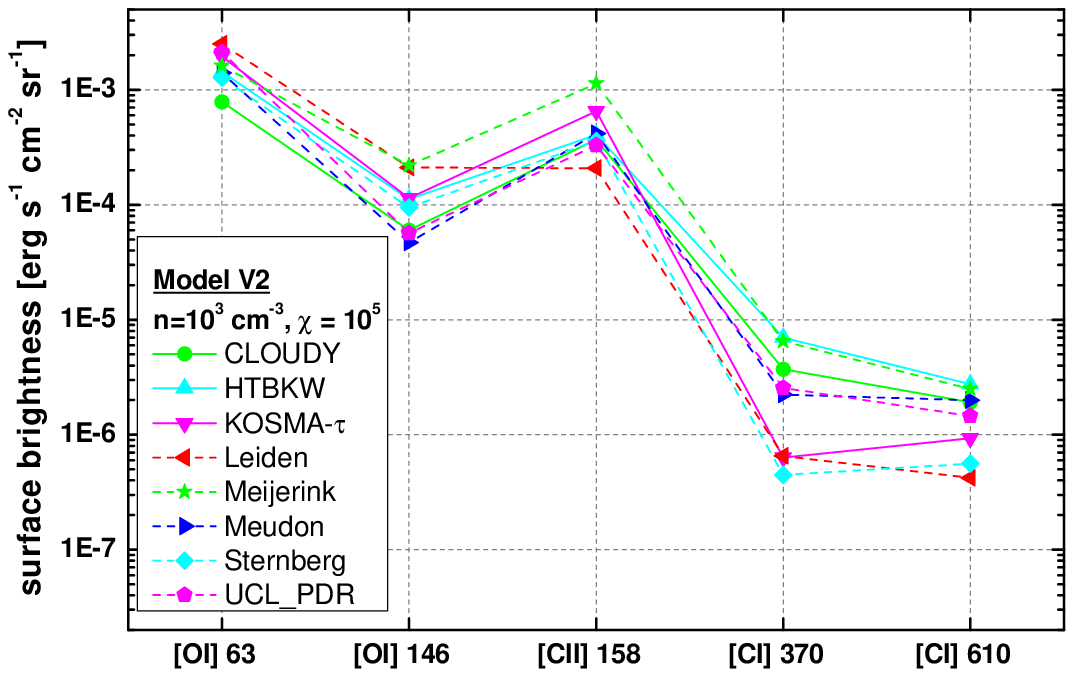}} 
 \caption{Model V2 (n=10$^{3}$~cm$^{-3}$, $\chi=10^5$): The plot shows the 
post-benchmark surface brightnesses of the main
 fine-structure cooling lines: [CII] 158 $\mu$m, [OI] 63, and 146 $\mu$m, and 
[CI] 610 and 370 $\mu$m.}\label{V2-surf}
 \end{figure}

\begin{figure*}
 \centering
 \includegraphics[width=16 cm]{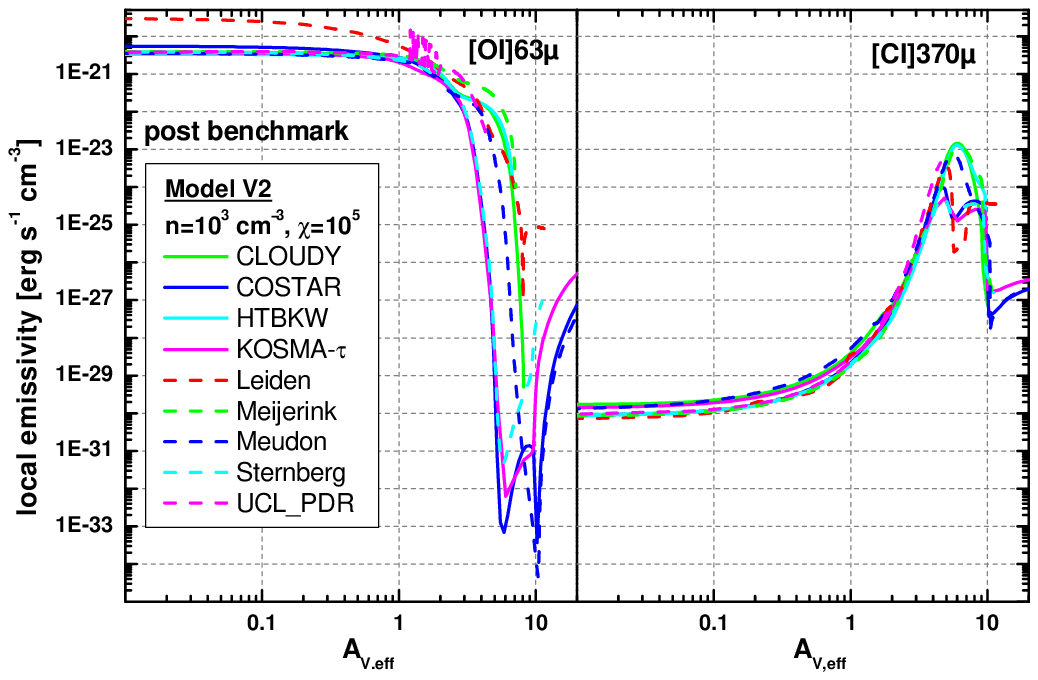}
 \caption{Model V2 (n=10$^{3}$~cm$^{-3}$, $\chi=10^5$): The post-benchmark local emissivities of
 [OI]~63$\mathrm{\mu m}$ (left column), and [CI]~310$\mathrm{\mu m}$ (right column). }\label{V2-OI-CI}
 \end{figure*}

In the benchmark models V1-V4 the various
 codes were  required to also solve the energy balance equations in order to derive the
 temperature structure of the model clouds. This of course introduces an additional source of
 variation between the codes. The chemical rate equations strongly depend on 
the local temperature, hence we expect
 a strong correlation between temperature differences and different chemical profiles of the model 
 codes. As a consequence of a differing density profile of e.g. CO and H$_2$ we also expect different
 shielding signatures. We will restrict ourselves to just a few exemplary non-isothermal results because
 a full analysis of the important non-isothermal models goes beyond the scope of this paper. To
 demonstrate the influence of
 a strong FUV irradiation we show results for the 
 benchmark model V2 with $n=10^{3}$~cm$^{-3}$, and $\chi=10^5$ in Figs.~\ref{V2-temp}-\ref{V2-OI-CI}. 
  The detailed treatment of the various
heating and cooling processes differs significantly from code to code. The only initial benchmark requirements
 was to treat the photoelectric (PE) heating according to \cite{BT94}. On one hand, 
this turned out to be not strict enough
 to achieve a sufficient agreement for the gas temperatures,
 on the other hand it was already too strict to be easily implemented for some codes, like {\tt Cloudy}, which 
calculates the PE heating self-consistently from a given dust composition. This 
demonstrates that there are limits to the degree of standardization. The calculation 
of the dust temperature was not standardized and varies from code to code. Since {\tt Lee96mod} only accounts
for constant temperatures, their model is not shown in the following plots. We only plot
 the final, post-benchmark status.
  
In Fig.~\ref{V2-temp} we show the gas temperature over $A_\mathrm{V,eff}$. 
 The general temperature profile is reproduced by all codes. Even so we note
considerable differences between different codes.
The derived temperatures at the surface
vary between $1600$ and $2500$~K.  For low values of $A_\mathrm{V,eff}$ the heating is dominated 
by PE heating due to the high FUV irradiation, and the main cooling is provided by [OI] 
and [CII] emission. It is interesting, that the dominant cooling line 
is the [OI]~63$\mathrm{\mu m}$ line (cf. Fig.~\ref{V2-OI-CI}, left plot),
 although its critical density is two orders of magnitude higher
than the local density ($n_\mathrm{cr}\approx 5\times10^5$~cm$^{-3}$). The highest surface temperature is 
calculated by {\tt Leiden}, while {\tt Meudon} computes the lowest temperature.  The bulk of models gives
surface temperatures near $1900$~K. All models qualitatively reproduce the temperature behavior at 
higher values of $A_\mathrm{V,eff}$ and show a minimum temperature of $~10$~K 
between $A_\mathrm{V,eff}\approx 5...10$, followed by a subsequent rise in temperature.
The only relevant heating contribution at $A_\mathrm{V,eff}>5$ comes from cosmic ray heating, which
hardly depends on $A_\mathrm{V,eff}$. 
 At $A_\mathrm{V,eff}>4$, the dominant cooling is by [CI] fine structure emission. This 
is a very efficient cooling process and the temperature reaches its minimum. At $A_\mathrm{V,eff}=10$ 
the atomic carbon density rapidly drops and CO cooling starts to exceed the fine structure 
cooling (cf. abundance profiles in Fig.~\ref{V2-big}). However, cooling by CO 
line emission is much less efficient, especially 
at these low total densities, and thus the temperature increases again.

 For the bulk of the cloud the heating contribution by H$_2$ vibrational deexcitation is 
negligible compared to photoelectric heating. Only {\tt Meijerink} and {\tt Leiden} predict 
comparable contributions from both processes. Unfortunately, the
exact treatment of this process was not standardized and depends very much on the 
detailed implementation (eg. the two-level approximation from \cite{burton90} or \cite{roellig05} 
vs. the solution of the full H$_2$ problem like in {\tt Meudon, Cloudy}, and {\tt Sternberg}). 
Generally the heating by H$_2$ vibrational deexcitation depends on the local density and the 
local mean FUV intensity, and should 
thus decrease at large values of $A_\mathrm{V,eff}$ and dominate the heating for denser clouds.

At $A_\mathrm{V,eff}\approx 2...3$ we
note a flattening  of the temperature curve in many models, followed by a 
steeper decline somewhat deeper inside the cloud.
This is not the case for {\tt HTBKW, KOSMA-$\tau$}, and {\tt Sternberg}. The reason for this difference is 
the [OI]~63$\mathrm{\mu m}$ cooling, showing a steeper decline for the three codes 
(Fig.~\ref{V2-OI-CI}, left plot).
For very large depths, {\tt KOSMA-$\tau$} produces slightly higher gas temperatures. This is
due to the larger dust temperature and the strongest H$_2$ vibrational deexcitation
heating at $A_\mathrm{V,eff}>10$.

In Fig.~\ref{V2-photo-H-both} we plot the photodissociation rate of H$_2$ (top left), the photoioniozation
rate of C (top right), and the density of H and H$_2$ over $A_\mathrm{V,eff}$ (bottom). {\tt Meudon's} 
 unshielded dissociation rate is by a factor three larger than the  median of 
$2.6\times 10^{-6}$~s$^{-1}$, and the {\tt Sternberg} value of $3.8\times 10^{-6}$~s$^{-1}$ 
is slightly larger for the same reason as discussed in section \ref{F1-F4}. 
The depth dependent shielding shows
good agreement between all codes, with slight variations. The different model geometry of 
{\tt KOSMA-$\tau$} is reflected in the slightly stronger shielding. {\tt Leiden} 
has the weakest shielding. Like some of the other codes (see Appendix), they account for the detailed 
H$_2$ problem when calculating the photodissociation rate, instead of applying tabulated shielding rates.
Yet these differences are small, since we are in a parameter regime ($\chi/n=100$), where 
the main shielding is
dominated by dust rather than by self shielding \citep{draine96}.
The density profiles
of H and H$_2$ are in good agreement. The stronger photodissociation in {\tt Meudon} is reflected in their
smaller H$_2$ density at the surface. All other H$_2$ densities correspond well to their dissociation
rates except for {\tt Cloudy}, which has a lower density at the surface without a corresponding 
photodissociation rate. This is a temperature effect. {\tt Cloudy} computes relatively low surface 
temperatures which lead to slightly lower H  densities at the surface. The central 
densities are also in good accord. The different H densities 
reflect the corresponding temperature profiles from Fig.~\ref{V2-temp}.

 The photoionization rate
of C is given in the top right plot in Fig.~\ref{V2-photo-H-both}. All models are in good 
agreement at the surface of the cloud. {\tt Meudon} and {\tt UCL\_PDR}
 drop slightly earlier than the bulk of the results. This is also reflected in their C density profiles
in Fig.~\ref{V2-big} (top right) which incline slightly earlier. Deep inside the cloud {\tt Sternberg} 
and {\tt HTBKW} show a steeper decline compared to the other codes.  
The agreement for the 
C$^+$ profile is also very good. At $A_\mathrm{V,eff}=5$ the densities drop by a factor of 10 and
remain constant until they drop at $A_\mathrm{V,eff}\approx 10$. This plateau is caused by the increase
in C density, compensating the FUV attenuating. {\tt Leiden's}
results show some deviations for $A_\mathrm{V,eff} > 10$. Their C density remains higher throughout the 
center, causing a slightly different carbon and 
oxygen chemistry at $A_\mathrm{V,eff}>10$. The calculated O and 
O$_2$ densities are given in  Fig.~\ref{V2-big} (bottom, right).
The dark cloud densities are in very good agreement among the models, with some deviations in the {\tt Leiden} 
values. The O$_2$ profiles show some variations between 
$A_\mathrm{V,eff}\approx 1$ and $10$ but these are minor deviations especially taking the fact that the densities
vary over 14 orders of magnitude from the outside to the center of the cloud! The differences in O$_2$
are also reflected in the CO plot (Fig.~\ref{V2-big}, bottom left). All codes produce very similar 
density profiles and dark cloud values. {\tt Leiden} gives a smaller CO density beyond 
$A_\mathrm{V,eff}=10$.  


In Fig.~\ref{V2-surf} we plot the total surface brightnesses of the main
 fine-structure cooling lines for the V2 model: [CII] 158 $\mu$m, [OI] 63, and 146 $\mu$m, and [CI] 610 and 370
 $\mu$m.
For the spherical PDR models, the surface brightness averaged over
the projected area of the clump is shown.
The surface brightness of these fine-structure lines is
smaller by typically a few 10\%, if calculated along
a pencil-beam toward the clump center as they are enhanced in the outer
cloud layers. Compared with the pre-benchmark results, 
the spread in $T_B$ has been decreased significantly from almost 3 orders of magnitude to a factor of 
3-5 for [CII] and [OI]. To explain the differences in Fig.~\ref{V2-surf} we plot
in Fig.~\ref{V2-OI-CI} the radial profiles of the local emissivities of
 [OI]~63$\mathrm{\mu m}$ and  [CI]~310$\mathrm{\mu m}$. {\tt Leiden} gives the highest 
[OI] brightnesses and
also computes higher local [OI] 63 $\mu$m emissivities for small values of $A_\mathrm{V,eff}$, 
shown in Fig.~\ref{V2-OI-CI}. {\tt COSTAR}, with very
similar results for the density profile and  
comparable gas temperatures, gives much smaller emissivities. The reason for these deviations is still unclear.
The model dependent spread in surface brightnesses is largest for the [CI] lines. {\tt HTBKW} computes 
10 times higher  line intensities for the [CI] 370$\mu$m transition than {\tt Sternberg}.
This can be explained as follows. Both codes show almost identical
 column densities and abundance profiles of C$^0$, yet the local emissivities are very different between
$A_\mathrm{V,eff}=4...9$ (Fig.~\ref{V2-OI-CI}). {\tt Sternberg}, together with some other codes, compute a local 
minimum for the cooling at $A_\mathrm{V,eff}\approx6$, while the {\tt HTBKW, Cloudy, Meijerink}, 
and {\tt Meudon} models
peak at the same depth. This can be explained as a pure temperature effect, since the codes
showing a [CI] peak compute a
significantly higher temperature at $A_\mathrm{V,eff}=6$: T({\tt HTBKW})=83~K, 
T({\tt Sternberg})=10~K. These different
temperatures at the C$^0$ abundance peak strongly influences the resulting [CI] surface 
brightnesses. Overall, the
model-dependent surface temperatures still vary significantly. This is due to the additional nonlinearity 
of the radiative transfer problem, which, under certain circumstances, amplifies even small 
abundance/temperature differences.                
 
\subsection{Review of participating codes}
It is not our intent to rate the various PDR model codes. Each  code was developed with a particular
field of application in mind and is capable to fulfill its developers expectations. The restrictions
artificially posed by the benchmark standards were additionally limiting the capacity 
of the participating model codes. 
Some models encountered for example  
major numerical difficulties in reaching a stable
temperature solution for the benchmark models V4, mainly caused by the requested H$_2$ formation
rate of $R=3\times10^{-18}T^{1/2}$~cm$^{3}$s$^{-1}$. This gives reasonable results 
for low temperatures, but diverges for
very high temperatures, resulting in an unreasonably high H$_2$ formation heating. 
Other codes also
 show similar numerical problems especially for the model V4. This numerical
noise vanishes when we apply more physically reasonable conditions. Nevertheless it was very instructive
to study the codes under these extreme conditions. 

Every participating code has its own strengths. The  {\tt Meudon} code and  {\tt Cloudy} are certainly the most 
complex codes in the benchmark, accounting for  most physical effects by explicit calculations,
starting from the detailed micro-physical processes,
making the least 
use of fitting formulae.

 {\tt Cloudy}  was originally developed to simulate extreme environments near
accreting black holes \citep{ferland88}. although it has been applied to
HII regions, planetary nebulae, and the ISM. \citet{ferland94} describe an 
early PDR calculation.  Its capabilities have been
greatly extended over the past several years \citep{vanhoof04,abel05,shaw05}. Due 
to the complexity of the code, it was initially not possible to turn off all 
implemented physical processes as required for the benchmark, but during this study they were able to 
adopt all benchmark requirements thus removing all major deviations. 

The codes  {\tt HTBKW},  {\tt Leiden},  {\tt Sternberg} and  {\tt KOSMA-$\tau$} are 
based on PDR models that began
their development 20 years ago
and have been supported and improved since then.  One of the main
differences between them is the model geometry and illumination. Plane-parallel geometry 
and uni-directional illumination
is assumed in {\tt HTBKW},  {\tt Leiden} and  {\tt Sternberg} and spherical geometry with an isotropically 
impinging FUV field in  {\tt KOSMA-$\tau$}. The chemistry adopted generally in {\tt HTBKW} is 
the smallest (46 species) compared with  {\tt Sternberg} (78) and  {\tt Leiden/KOSMA-$\tau$} 
(variable).  {\tt Leiden, Sternberg and KOSMA-$\tau$} 
explicitly solve the H$_2$ problem (full ro-vib level population) and determine 
the corresponding shielding by integrating all absorption coefficients  
while  {\tt HTBKW} uses shielding 
functions and a single-line approximation for H$_2$. {\tt Cloudy} is also  capable of explicitly
calculating a fully (v,J) resolved H$_2$ model, but this capability was switched
off in the final model. Instead they used a 3-level approximation there.  {\tt Leiden} 
and  {\tt Meudon} are the only codes in the benchmark
explicitly calculating the CO shielding, all other codes use shielding factors.  {\tt HTBKW} is 
additionally accounting for X ray and PAH heating 
and computes a large number of observational line intensities, while  {\tt Leiden} focuses on the
 line emission from the
  main PDR coolants C$^+$, C, O, and CO. However it is possible to couple their PDR output with a more
sophisticated radiative transfer code such as RATRAN \citep{ratran} to calculate emission
 lines. This is also done
by {\tt KOSMA-$\tau$}, using ONION \citep{gierens92} or SimLine \citep{simline}.  {\tt COSTAR} was developed 
in order to model circumstellar disks, featuring any given disk density profile in radial direction 
and scale height in vertical direction. It uses uni-directional FUV illumination and can treat a surrounding 
isotropic interstellar FUV field in addition to the uni-directional stellar field. It 
computes a relatively small chemical network (48 species) but also accounts for 
freeze-out onto grains and desorption
effects. It relies on shielding functions for H$_2$ and CO and does not calculate observational 
line intensities up to now. Nevertheless most of the  {\tt COSTAR} results are in good agreement with the 
other code results for most of the benchmark models, demonstrating that it correctly accounts for the
important PDR physics and chemistry.  {\tt UCL\_PDR} is a plane-parallel model focused 
on time-dependent chemistries with freeze-out and desorption. Its main features 
are a fully time-dependent treatment - including time-varying density and  
radiation profiles - and its speed, which makes it suitable for  
parameter search studies where a large number of models need to be  
run. It can also be coupled with the SMMOL radiative transfer code  
\citep{rawlings01} for a detailed treatment  
of the PDR emission properties.
{\tt Lee96mod} was developed from the time-dependent chemical model by Lee, Herbst, and collaborators. It is
strongly geared toward complex chemical calculations and only accounts for constant temperatures, neglecting
local cooling and heating. {\tt Meijerink} is a relatively 
young model with special emphasis on XDRs (X-ray dominated regions) which 
quickly evolved in the course of this study and we refer
to \citet{meijerink05} for a detailed review of the current status. In the Appendix we 
give a tabular overview of all main model characteristics.

\section{Concluding remarks}\label{summary}
We present the latest result in a community wide comparative study among PDR model codes. PDR
models are available for almost 30 years now and are established as a common and 
trusted tool for the interpretation of observational data.
The PDR model experts and code-developers have long recognized that 
the existing codes may deviate significantly in their results,
so that observers must not blindly use the output from one of
the codes to interpret line observations. The PDR-benchmarking workshop was a first attempt to solve this
problem by separating numerical and conceptional differences in the codes, and removing
ordinary bugs so that the PDR codes finally turn into a reliable tool for the interpretation
of observational data.

Due to their complex nature
it is not always straightforward to compare results from different PDR models with each
other. Given the large number of input paramters,  it is usually possible to derive more 
than one set of physical parameters by comparing observations with model predictions, especially
when one is chiefly interested in mean densities and temperatures. 
 Our goal was to understand the mutual differences in the different model
results and to work toward a better understanding of the key processes involved in PDR
modeling. The comparison has revealed the importance of an accurate treatment of various processes,
which require further studies.

The workshop and the following benchmarking activities were
 a success regardless of many open issues. The major 
results of this study are:
\begin{itemize}
\item The collected results from all participating models represent an excellent
reference for all present PDR codes and for those to be developed in the future. For the first 
time such a reference is easily available not only in graphical form  but also as raw data. (URL:
{\tt \small http://www.ph1.uni-koeln.de/pdr-comparison})
\item We present an overview of the common PDR model codes and summarize their
properties and field of application
\item As a natural result all participating PDR codes are now better debugged, much better understood,  
and many differences between the results from different groups are now much clearer
resulting in good guidance for further improvements. 
\item Many critical parameters, model properties and physical processes have been identified
or better understood in the course of this study.
\item We were able to increase the agreement in model prediction for all benchmark models. 
Uncertainties still remain, visible e.g. in the deviating temperature profiles
of model V2 (Fig.~\ref{V2-temp}) or the large differences
for the H$_2$ photo-rates and density profiles in model V4 (cf. online data archive).
\item All PDR models are heavily dependent on the chemistry and micro-physics involved in PDRs. Consequently
the results from PDR models are only as reliable as the description of the 
microphysics (rate coefficients, etc.) they are based on.
\end{itemize}
One of the lessons from this study is that observers should not take the PDR results too literally
to constrain, for example, physical parameters like density and radiation field in the region they 
observe. The current benchmarking shows that all trends are consistent between codes
but that there remain differences in absolute values of observables. Moreover it is
not possible to simply infer how detailed differences in density or temperature translate into differences
in observables. They are the result of a complex, nonlinear interplay between density, temperature, and
radiative transfer.
We want to emphasize again, that all participating PDR codes are much 'smarter' than required
during the benchmark. Many sophisticated model features have been switched off in order to provide
comparable results. Our intention was technical not physical. The presented
results are not meant to model any real astronomical object and should not be applied as such to any 
such analysis. The current benchmarking results are not meant as our recommended or best values,
but simply as a comparison test.
During this study we demonstrated, that an increasing level of standardization results in a significant
reduction of the model dependent scatter in PDR model predictions. It is encouraging to note the 
overall agreement in model results. On the other hand it is important to understand that small changes
may make a big difference. We were able to identify a number of these key points, e.g. the influence of
excited hydrogen, or the importance of secondary photons induced by cosmic rays.

Future work should focus on the energy balance problem, clearly evident from the
sometimes significant scatter in the results for the non-isothermal models V1-V4.
The heating by photoelectric emission is closely related
to the electron density and to the detailed description of grain charges, grain surface
recombinations and photoelectric yield. The high temperature regime also requires an enlarged
set of cooling processes. Another important consideration to be adressed, especially when it
comes to comparisons with observations is the model density structure, i.e. clumping or
gradients.  As a consequence we plan to continue our benchmark effort 
in the future. This should include a calibration on real observational findings as well. 

\citeindextrue
\begin{acknowledgements}
We thank the Lorentz Center, Leiden, for hosting the workshop and for the perfect 
organization, supplying a very productive environment. The workshop and this work was 
partly funded by the Deutsche Forschungs Gesellschaft DFG via Grant SFB494 and by a 
Spinoza grant from the Netherlands Organization for Scientific Research (NWO). We also 
would like to thank the referee and the editor for making helpful suggestions which helped to
  improve the manuscript. 
\end{acknowledgements}

\appendix
\section{Characteristics of Participating PDR Codes }
In Tab.~\ref{characteristics} we summarize the most important characteristics of the participating
PDR codes. This table summarizes the full capabilities of the PDR codes
and is not limited to the benchmark standards. It has been extracted from detailed characteristics 
sheets, available online for 
all codes: {\small \tt http://www.ph1.uni-koeln.de/pdr-comparison}. 
\clearpage
\onecolumn
% Table generated by Excel2LaTeX from sheet 'Tabelle1 (2)'
\begin{longtable}{ll|c|c|c|c|c|c|c|c|c|c|c}
\caption{Full capabilities of the PDR model codes participating in the Leiden comparison study}\label{characteristics}\\
\hline \hline
    {\bf } &     {\it } &\rotatebox{90}{\bf Cloudy} & \rotatebox{90}{\bf COSTAR} & \rotatebox{90}{\bf Meudon} & \rotatebox{90}{\bf UCL\_PDR}  & \rotatebox{90}{\bf HTBKW} & \rotatebox{90}{\bf KOSMA-$\tau$} & \rotatebox{90}{\bf Aikawa} & \rotatebox{90}{\bf Leiden} & \rotatebox{90}{\bf Lee96mod} & \rotatebox{90}{\bf Sternberg} & \rotatebox{90}{\bf Meijerink} \\ \hline 
\endfirsthead
\caption{continued.}\\
\hline \hline
    {\bf } &     {\it } &{\rotatebox{90}{\bf Cloudy}} & \rotatebox{90}{\bf COSTAR} & \rotatebox{90}{\bf Meudon} & \rotatebox{90}{\bf UCL\_PDR} &  \rotatebox{90}{\bf HTBKW} & \rotatebox{90}{\bf KOSMA-$\tau$} & \rotatebox{90}{\bf Aikawa} & \rotatebox{90}{\bf Leiden} & \rotatebox{90}{\bf Lee96mod} & \rotatebox{90}{\bf Sternberg} & \rotatebox{90}{\bf Meijerink} \\ \hline 
\endhead 
\multicolumn{13}{c}{\rule[-3mm]{0mm}{8mm}\bf GEOMETRY} \\ 
\hline 
    {\bf } & { spherical} &          x &            &            &            &                 &          x &            &            &            &            &            \\ \hline

    {\bf } & { plane-parallel, finite } &          x &            &          x &            &            x &            &            &          x &            &            &            \\ \hline

    {\bf } & { plane-parallel, semi-infinite } &          x &            &          x &          x &              x &            &          x &          x &          x &          x &          x \\ \hline

    {\bf } & { circumstellar disc} &          x &          x &            &            &                &            &            &         x   &            &            &            \\ \hline

    {\bf } & { ensemble of clouds} &            &            &            &            &               &          x &            &            &            &            &            \\ \hline
\multicolumn{13}{c}{\rule[-3mm]{0mm}{8mm}\bf DENSITY}    \\
\hline 
    {\bf } & { homogeneous} &          x &          x &          x &          x &              x &          x &          x &          x &          x &          x &          x \\ \hline

    {\bf } & { density-law} &          x &       x     &          x &          x &               &          x &          x &          x &          x &          x &            \\ \hline

    {\bf } & {  time dependent} &          x &            &            &         x   &                  &            &            &            &            &            &            \\ \hline

    {\bf } & { velocity field}  &          x &            &            &            &                &          x &            &            &            &            &            \\ \hline

    {\bf } & {\it \hspace{2cm}v = const} &          x &            &            &            &                &          x &            &            &            &            &            \\ \hline

    {\bf } & {\it  \hspace{2cm}v= v(r,...)} &         x   &            &            &            &                 &            &            &            &            &            &            \\ \hline
\multicolumn{13}{c}{\rule[-3mm]{0mm}{8mm}\bf RADIATION}      \\ 
\hline
    {\bf } & { isotropic radiation field} &            &            &          x &            &               &          x &            &            &            &            &            \\ \hline

    {\bf } & { uni-directional radiation field} &          x &          x &          x &          x &              x &            &          x &          x &          x &          x &          x \\ \hline

    {\bf } & { combination of isotropic+illuminating star} &            &            &          x &            &                   &            &            &            &            &            &            \\ \hline

    {\bf } &     {\it } &            &            &            &            &                &            &            &            &            &            &            \\ \hline

    {\bf } & { Habing field} &          x &            &            &          x &              x &            &            &          x &            &            &          x \\ \hline

    {\bf } & { Draine field} &          x &          x &          x &            &               x &          x &            &          x &            &          x &            \\ \hline

    {\bf } & { optional star} &          x &            &          x &            &                 &            &            &         x   &            &            &            \\ \hline

    {\bf } & { detailed SED} &          x &            &          x &            &                 &            &            &            &            &            &            \\ \hline

    {\bf } & { other} &         x   &            &            &            &                &            &          x &          x &          x &            &            \\ \hline

    {\bf } &     {\it } &            &            &            &            &                &            &            &            &            &            &            \\ \hline

    {\bf } & { external radiation source} &          x &          x &          x &          x &              x &          x &          x &          x &          x &          x &          x \\ \hline

    {\bf } & { internal radiation source} &            &            &            &            &                &            &            &            &            &            &            \\ \hline
\multicolumn{13}{c}{\rule[-3mm]{0mm}{8mm}\bf CHEMISTRY}        \\ 
\hline
    {\bf } & { stationary chemistry} &          x &          x &          x &            &               x &          x &            &          x &            &          x &          x \\ \hline

    {\bf } & { time-dependent chemistry} &          x &            &            &          x &                  &            &          x &            &          x &            &            \\ \hline

    {\bf } & { advection flow} &         x   &            &            &            &               &            &            &            &            &            &            \\ \hline

    {\bf } &     {\it } &            &            &            &            &              &            &            &            &            &            &            \\ \hline

    {\bf } & { UMIST95} &         x   &            &          x &          x &               &         x   &            &          x &            &          x &          x \\ \hline

    {\bf } & { UMIST99} &         x   &            &            &          x  &              x &          x &            &            &            &         x  &  x          \\ \hline

    {\bf } &  { NSM} &            &            &          x &            &            &            &          x &            &          x &            &            \\ \hline

    {\bf } & { other database} &          x &          x &          x &            &                 &          x &            &          x &            &          x &            \\ \hline

    {\bf } &     {\it } &            &            &            &            &                  &            &            &            &            &            &            \\ \hline

    {\bf } & { fixed number of species} &          x &          x &            &          x &              x &            &          x &            &          x &          x &            \\ \hline

    {\bf } & { variable number of species} &            &            &          x &            &                &          x &            &          x &            &            &          x \\ \hline

    {\bf } & {\it  \hspace{2cm}number of species} &         96 &         48 &            &        128 &             46 &            &        577 &            &        419 &         78 &            \\ \hline

    {\bf } &     {\it } &            &            &            &            &                &            &            &            &            &            &            \\ \hline

    {\bf } & { PAH's included} &          x &            &          x &           &             x &          x &            &          x &            &            &          x \\ \hline

    {\bf } &     {\it } &            &            &            &            &                &            &            &            &            &            &            \\ \hline

    {\bf } & { freeze-out on grains included} &          x &          x &          x &          x &                 &            &          x &            &            &            &            \\ \hline

    {\bf } &     {\it } &            &            &            &            &                &            &            &            &            &            &            \\ \hline

    {\bf } &  H$_2${ formation on grains} &          x &          x &          x &          x &                x &          x &          x &          x &          x &          x &          x \\ \hline

    {\bf } & { formation of other molecules on grains} &            &            &          x &          x &                  &            &          x &            &            &            &            \\ \hline

    {\bf } &     {\it } &            &            &            &            &               &            &            &            &            &            &            \\ \hline

    {\bf } & { desorption mechanisms included} &            &          x &          x &            &                &            &          x &            &            &            &            \\ \hline

    {\bf } & {\it  \hspace{2cm}thermal desorption} &         x   &          x &            &            &                &            &        x    &            &            &            &            \\ \hline

    {\bf } & {\it  \hspace{2cm}photoevaporation} &            &            &          x &            &                 &            &            &            &            &            &            \\ \hline

    {\bf } & {\it  \hspace{2cm}CR spot heating} &         x   &            &          x &            &                &            &          x &            &            &            &            \\ \hline

    {\bf } & {\it  \hspace{2cm}grain-grain collisions} &            &            &          x &            &                  &            &            &            &            &            &            \\ \hline

    {\bf } &     {\it } &            &            &            &            &                 &            &            &            &            &            &            \\ \hline

    {\bf } & { metallicity included} &          x &          x &          x &          x &            x &          x &            &          x &            &          x &          x  \\ \hline
\multicolumn{13}{c}{\rule[-3mm]{0mm}{8mm}\bf ISOTOPOMERS}      \\
\hline
    {\bf } &  $^{13}$C &          x &            &          x &            &             &          x &            &          x &            &            &          x  \\ \hline

    {\bf } &   $^{17}$O &            &            &            &            &               &            &            &            &            &            &            \\ \hline

    {\bf } &  $^{18}$O &            &            &          x &            &                &          x &            &          x &            &            &            \\ \hline

    {\bf } &     D &          x &            &          x &            &                  &            &          x &          x &            &            &            \\ \hline

    {\bf } &     {\it } &            &            &            &            &                    &            &            &            &            &            &            \\ \hline
\multicolumn{13}{c}{\rule[-3mm]{0mm}{8mm}\bf THERMAL BALANCE}       \\
\hline
    {\bf } & { fixed temperature} &          x &         x   &          x &          x  &              &          x &            &          x  &          x &          x &          x  \\ \hline

    {\bf } & { temperature determined from energy balance} &          x &          x &          x &          x &              x &          x &          x &          x &            &          x &          x \\ \hline
\multicolumn{13}{c}{\rule[-3mm]{0mm}{8mm}\bf COOLING}        \\ 
\hline
    {\bf } & { gas-grain cooling} &          x &          x &          x &          x &             x &          x &          x &          x &            &          x &          x \\ \hline

    {\bf } & { radiative recombination} &          x &            &          x  &            &                 &          x &            &            &            &          x &            \\ \hline

    {\bf } & { chemical balance} &            &            &          x &            &                &            &            &            &            &            &            \\ \hline

    {\bf } &     {\it } &            &            &            &            &                 &            &            &            &            &            &            \\ \hline

    {\bf } & [OI]{  lines} &          x &          x &          x &          x &             x &          x &          x &          x &            &          x &          x \\ \hline

    {\bf } & $^{12}$CO{  rotational lines} &          x &          x &          x &          x &             x &          x &          x &          x &            &            &          x \\ \hline

    {\bf } & $^{13}$CO{ rotational lines} &          x &            &          x &            &                &          x &          x &            &            &            &          x \\ \hline

    {\bf } &  [CII]{ line} &          x &          x &          x &          x &             x &          x &          x &          x &            &          x &          x \\ \hline

    {\bf } & [CI]{  lines} &          x &          x &          x &          x &             x &          x &          x &          x &            &          x &          x \\ \hline

    {\bf } &  [SiII] { lines} &          x &            &          x &            &              x &          x &            &            &            &          x &          x \\ \hline

    {\bf } &  OH  { rotational lines} &            &            &          x &            &               x &          x &            &            &            &          x  &          x \\ \hline

    {\bf } &  H$_2$O  { rotational lines} &            &            &          x &            &             x &          x &            &            &            &         x   &          x \\ \hline

    {\bf } &  H$_2$ { rotational lines} &          x &            &          x &          x &             &            &            &            &            &         x   &          x  \\ \hline

    {\bf } &  HD{ rotational lines} &          x &            &          x &            &                 &            &            &            &            &            &            \\ \hline

    {\bf } &  [OI] 6300$\AA$ {metastable lines} &          x &          x &         x   &          x &               &            &            &         x   &            &         x   &          x  \\ \hline

    {\bf } &  CH { rotational lines} &            &          x &            &            &               &            &            &            &            &          x  &            \\ \hline

    {\bf } & { Ly $\alpha$ metastable lines} &          x &          x &            &          x &               &            &            &          x  &            &            &          x  \\ \hline

    {\bf } &  Fe(24$\mu$,34$\mu$), [FeII](26$\mu$,35.4$\mu$) &          x &            &            &            &                x &            &            &            &            &          x  &         x   \\ \hline

    {\bf } &  H$_2$ { (rot-vib)} &          x &            &          x  &            &            x &            &            &            &            &          x  &          x  \\ \hline
\multicolumn{13}{c}{\rule[-3mm]{0mm}{8mm}\bf HEATING}   \\ 
\hline
    {\bf } &  H$_2$ { vibrational deexcitation} &          x &          x &          x &          x &             x &          x &          x &          x &            &          x &          x \\ \hline

    {\bf } & {\it  \hspace{2cm}single line approx.} &          x &          x &            &          x &              x &            &          x &          x &            &            &          x \\ \hline

    {\bf } & {\it \hspace{2cm}only n-levels, but no J} &            &            &            &            &                 &          x &            &            &            &          x &            \\ \hline

    {\bf } & {\it  \hspace{2cm}full rot-vib treatment} &          x &            &          x &            &                 &            &            &            &            &          x &            \\ \hline

    {\bf } &  H$_2${ dissociation} &          x &          x &          x &          x &             x &          x &          x &          x &            &          x &          x \\ \hline

    {\bf } &  H$_2${ formation} &          x &          x &          x &          x &              x &          x &            &          x &            &          x &          x  \\ \hline

    {\bf } & { CR heating} &          x &          x &          x &          x &          x &          x &          x &          x &            &          x &          x \\ \hline

    {\bf } & { PE heating} &          x &          x &          x &          x &            x &          x &          x &          x &            &          x &         x   \\ \hline

    {\bf } & { XR heating} &          x &            &            &          x &           x &            &            &            &            &            &         x   \\ \hline

    {\bf } & { PAH heating} &          x &            &          x &          x &            x &          x &            &          x &            &            &        x    \\ \hline

    {\bf } & { photoionization} &          x &          x &          x  &          x &             &            &            &          x &            &            &          x \\ \hline

    {\bf } & {\it  \hspace{2cm}carbon ionization heating} &          x &          x &         x   &          x &               &            &            &          x &            &         x   &         x   \\ \hline

    {\bf } & {\it  \hspace{2cm}other species (Si, etc.)} &          x &            &          x  &            &                  &            &            &            &            &            &            \\ \hline

    {\bf } & { gas-grain collisions} &          x &         x   &         x   &          x  &              &            &            &          x  &            &          x  &         x   \\ \hline

    {\bf } & { turbulence heating} &          x &            &         x   &          x &               &            &            &            &            &            &            \\ \hline

    {\bf } & { chemical balance} &            &            &          x &          x  &               &            &            &          x &            &            &            \\ \hline
\multicolumn{13}{c}{\rule[-3mm]{0mm}{8mm}\bf UV TRANSFER}      \\ 
\hline
    {\bf } & { solved self-consistently} &          x &          x &          x &          x &             x &          x &            &          x &            &          x &          x \\ \hline

    {\bf } & { simple exponential attenuation} &          x &          x &          x &          x &              x &          x &          x &          x &          x &          x  &          x \\ \hline

    {\bf } & { bi-exponential attenuation} &            &          x &            &            &                &            &            &            &            &          x &            \\ \hline

    {\bf } & { full RT in lines} &            &            &          x &            &            &            &            &        x    &            &            &            \\ \hline
\multicolumn{13}{c}{\rule[-3mm]{0mm}{8mm}\bf DUST}  \\ 
\hline
    {\bf } & { treatment of rad. transfer} &          x &            &          x &            &             x &          x &            &          x &            &          x &         x   \\ \hline

    {\bf } & { grain size distribution} &          x &            &          x &          x &              &          x &            &            &            &           &            \\ \hline

    {\bf } & { extinction/scattering law} &          x &          x &          x &          x &             x &          x &          x &          x &          x &          x &            \\ \hline

    {\bf } & { albedo} &          x &            &          x &          x &               &            &            &          x &            &            &         x   \\ \hline

    {\bf } & { scattering law} &          x &            &          x &           &              &            &            &          x &            &            &            \\ \hline
\multicolumn{13}{c}{\rule[-3mm]{0mm}{8mm}\bf H$_2$ SHIELDING}     \\
\hline
    {\bf } & { shielding factors} &          x &          x &            &            &             x &            &          x &            &          x &         x   &          x  \\ \hline

    {\bf } & { single line} &          x &            &            &          x &              &            &            &            &            &            &          x \\ \hline

    {\bf } & { detailed solution} &          x &            &          x &            &            &          x &            &          x &            &           &            \\ \hline
\multicolumn{13}{c}{\rule[-3mm]{0mm}{8mm}\bf CO SHIELDING}   \\ 
\hline
    {\bf } & { shielding factors} &         x   &          x &            &          x &            x &          x &          x &          x &          x &          x &          x  \\ \hline

    {\bf } & { single line} &          x &            &            &            &               &            &            &          x &            &            &          x \\ \hline

    {\bf } & { detailed solution} &            &            &          x &            &                &            &            &          x &            &            &            \\ \hline

    {\bf } & { isotope selective photodissociation} &            &            &          x &            &               &          x &            &          x &            &            &          x  \\ \hline
\multicolumn{13}{c}{\rule[-3mm]{0mm}{8mm}\bf UV PROFILE FUNCTION}   \\ 
\hline
    {\bf } & { Gaussian} &            &            &            &          x &           x &            &            &            &            &            &            \\ \hline

    {\bf } & { Voigt} &          x &            &          x &            &              &            &            &          x &            &          x &          x  \\ \hline

    {\bf } &  { Box} &            &            &            &            &              &            &            &            &            &            &            \\ \hline

    {\bf } & { other} &            &            &            &            &               &            &            &            &            &            &            \\ \hline
\multicolumn{13}{c}{\rule[-3mm]{0mm}{8mm}\bf RADIATIVE TRANSFER IN COOLING LINES}  \\
\hline
    {\bf } & { escape probability} &          x &          x &          x &          x &             x &          x &          x &          x &            &          x &          x \\ \hline

    {\bf } & { other} &            &            &            &            &                &            &            &            &            &            &            \\ \hline

    {\bf } & { IR pumping} &          x &          x &            &           &              x &            &            &          x &            &            &          x  \\ \hline
\multicolumn{13}{c}{\rule[-3mm]{0mm}{8mm}\bf OBSERVATIONAL LINES}  \\ 
\hline
    {\bf } & { self-consistent treatment with cooling} &          x &            &            &          x &                 &            &            &            &            &            &            \\ \hline

    {\bf } & { escape probability} &          x &            &            &          x  &              &      x      &          x &          x &            &          x &          x \\ \hline

    {\bf } & { other} &            &            &          x &            &            &          x &            &            &            &            &            \\ \hline

    {\bf } &  { H$_2$} &          x &            &          x &            &               &            &            &          x  &            &          x  &            \\ \hline

    {\bf } &   { HD} &            &            &          x &            &                &            &            &           x &            &          x  &            \\ \hline

    {\bf } & { $^{12}$CO  } &          x &            &          x &          x &              x &          x &          x &            &            &            &          x  \\ \hline

    {\bf } & { $^{13}$CO} &          x &            &          x &            &                 &          x &            &            &            &            &           x \\ \hline

    {\bf } & { C$^{18}$O} &            &            &          x &            &                &          x &            &            &            &            &            \\ \hline

    {\bf } & { $^{13}$C$^{18}$O} &            &            &          x &            &              &          x &            &            &            &            &            \\ \hline

    {\bf } &    { [OI]} &          x &            &          x &          x &              x &          x &          x &          x &            &          x &          x \\ \hline

    {\bf } &   { [CII]} &          x &            &          x &          x &               x &          x &          x &          x &            &          x &          x \\ \hline

    {\bf } &   { [CI]} &          x &            &          x &          x &             x &          x &          x &          x &            &          x &          x \\ \hline

    {\bf } & { Si$^+$   } &          x &            &          x &            &          x &            &            &            &            &            &          x  \\ \hline

    {\bf } &   { CS} &            &            &          x &            &              &            &            &            &            &            &          x  \\ \hline

    {\bf } & { H$_2$O  } &            &            &            &            &             x &            &            &            &            &            &            \\ \hline

    {\bf } & { H$_2^{18}$O} &            &            &            &            &               &            &            &            &            &            &            \\ \hline

    {\bf } & { HCO$^+$} &            &            &          x &            &              x &         x  &            &            &            &            &         x   \\ \hline

    {\bf } &   { OH} &            &            &            &            &               x &            &            &            &            &            &            \\ \hline

    {\bf } &  { [SiI]} &          x &            &            &            &              x &         x   &            &            &            &            &            \\ \hline

    {\bf } & { [SI],[SII]} &          x &            &            &            &               x &          x  &            &            &            &            &          x  \\ \hline

    {\bf } & { [FeI], [FeII]} &          x &            &            &            &           x &            &            &            &            &            &        x    \\ \hline
\multicolumn{13}{c}{\rule[-3mm]{0mm}{8mm}\bf COMPUTED LINE PROPERTIES}\\ 
\hline
    {\bf } & { resolved line profile} &            &            &          x &            &            x &          x &            &          x &            &            &            \\ \hline

    {\bf } & { continuum rad./rad transfer in UV} &          x &            &          x &            &                &            &            &            &            &            &            \\ 

    {\bf } & { line center intensities} &          x &            &          x &            &              &          x &            &          x  &            &            &            \\ \hline

    {\bf } & { line integrated intensities} &          x &            &          x &          x &             x &          x &            &          x &            &            &          x \\ \hline

    {\bf } & { optical depths} &          x &            &          x &          x &              x &          x &            &          x &            &            &          x  \\ \hline

    {\bf } & { Gaussian line profile} &          x &            &          x &            &              x &          x &            &          x &            &            &          x  \\ \hline

    {\bf } & { box line profile} &            &            &            &            &                &            &            &            &            &            &            \\ \hline

    {\bf } & { turbulence included} &          x &            &          x &            &               &          x &            &            &            &            &          x  \\ \hline
\multicolumn{13}{c}{\rule[-3mm]{0mm}{8mm}\bf COLLISIONS} \\ 
\hline
    {\bf } &  { H-H} &          x &            &            &            &            x &            &            &            &            &          x &            \\ \hline

    {\bf } & { H$_2$-H} &          x &            &          x &          x &              x &            &            &          x &            &          x &         x   \\ \hline

    {\bf } & { H$_2$ - H+} &          x &            &          x &            &                &            &            &            &            &          x &            \\ \hline

    {\bf } & { H$_2$ - e} &          x &            &         x   &            &                x &            &            &            &            &          x &            \\ \hline

    {\bf } & { H$_2$ - H$_2$} &          x &            &          x &            &               x &            &            &          x &            &          x &         x   \\ \hline

    {\bf } & { CO-H} &          x &          x &          x &          x &              x &            &            &          x &            &            &          x \\ \hline

    {\bf } & { CO-H$_2$} &          x &          x &         x   &          x &             x &          x &            &          x &            &            &          x \\ \hline

    {\bf } & { CO-e} &          x &          x &            &          x &                 &          x &            &            &            &            &          x \\ \hline

    {\bf } & { CO - He} &            &            &          x &          x &                &            &            &            &            &            &         x   \\ \hline

    {\bf } &  { C-H} &         x   &          x &        x    &          x &            x &          x &            &          x &            &          x &          x  \\ \hline

    {\bf } & { C-H$_2$} &         x   &          x &         x   &          x &              x &          x &            &          x &            &            &        x    \\ \hline

    {\bf } &  { C-e} &          x &            &         x   &          x &               &            &            &            &            &            &          x  \\ \hline

    {\bf } & { C - He} &         x   &            &         x   &          x &                &            &            &            &            &            &            \\ \hline

    {\bf } & { C - H$_2$O} &           &            &            &            &                &            &            &            &            &            &            \\ \hline

    {\bf } & { C$^+$ - H} &          x &          x &          x &          x &              x &            &            &          x &            &            &         x   \\ \hline

    {\bf } & { C$^+$ - H$_2$} &          x &          x &          x &          x &            x &          x &            &          x &            &            &          x \\ \hline

    {\bf } & { C$^+$ - e} &          x &          x &            &          x &            x &          x &            &          x &            &            &         x   \\ \hline

    {\bf } & { O - H} &          x &          x &          x &          x &            x &          x &            &          x &            &          x &          x  \\ \hline

    {\bf } & { O - H$_2$} &          x &          x &          x &          x &             x &          x &            &          x &            &          x &          x \\ \hline

    {\bf } & { O - H+} &          x &            &         x   &          x &              &            &            &          x &            &            &            \\ \hline

    {\bf } & { O - e} &          x &          x &         x   &          x &              x &            &            &            &            &            &            \\ \hline

    {\bf } & { O - He} &          x &            &         x   &          x &               &            &            &            &            &            &            \\ \hline

    {\bf } & { OH - H} &           &            &            &            &              &            &            &            &            &            &            \\ \hline

    {\bf } & { OH - He} &            &            &            &            &               &            &            &            &            &            &            \\ \hline

    {\bf } & { OH - H$_2$} &           &            &            &            &              x &          x &            &            &            &            &            \\ \hline

    {\bf } & { H$^-$ - H} &          x &            &            &            &              &            &            &            &            &            &            \\ \hline

    {\bf } & { H$_2$O - e} &           &            &            &            &              &            &            &            &            &            &            \\ \hline

    {\bf } & { H$_2$O - H} &           &            &            &            &               &            &            &            &            &            &         x   \\ \hline

    {\bf } & { H$_2$O - H$_2$} &            &            &            &            &           x &            &            &            &            &            &        x    \\ \hline

    {\bf } & { H$_2$O - O} &           &            &            &            &             &            &            &            &            &            &            \\ \hline

    {\bf } & { dust - H/H$_2$} &          x &            &            &            &             x &            &            &            &            &            &            \\ \hline

    {\bf } & { dust-any} &          x &            &            &            &               &            &            &            &            &            &            \\ \hline

    {\bf } & { Si$^+$ - H} &          x &            &          x &            &               &         x   &            &            &            &            &            \\ \hline

    {\bf } & { HD - H} &            &            &          x &            &              &            &            &            &            &            &            \\ \hline

    {\bf } & { HD - H$_2$} &            &            &          x &            &               &            &            &            &            &            &            \\ \hline

    {\bf } & { PAH-any} &          x &            &            &            &             x &            &            &            &            &            &            \\ \hline
\multicolumn{13}{c}{\rule[-3mm]{0mm}{8mm}\bf OUTPUT} \\
\hline
    {\bf } & { abundance profile over (A$_V$/depth)} &          x &          x &          x &          x &            x &          x &          x &          x &          x &          x &          x \\ \hline

    {\bf } & { column density over (A$_V$/depth)} &          x &         x   &          x &         x   &                 &          x &            &            &            &          x &         x   \\ \hline

    {\bf } & { temperature profile over (A$_V$/depth)} &          x &          x &          x &          x &              x &          x &          x &          x &            &          x &          x \\ \hline

    {\bf } & { emitted intensities} &          x &            &          x &          x &           x &          x &            &          x &            &          x  &          x \\ \hline

    {\bf } & { opacities at line center} &            &            &          x &            &            x &          x &            &          x &            &         x   &         x   \\ \hline

    {\bf } & { heating and cooling rates over (A$_V$/depth)} &          x &            &          x &          x  &             &          x &            &         x   &            &          x &         x   \\ \hline

    {\bf } & { chemical rates over (A$_V$/depth)} &            &            &          x &         x   &               &          x &            &         x   &            &          x &          x  \\ \hline

    {\bf } & { excitation diagram of }H$_2$ &          x &            &          x &            &                &            &            &         x   &            &          x  &            \\ \hline

\end{longtable}

\end{document}